\documentclass[onecolumn,12pt,draftclsnofoot]{IEEEtran}

\usepackage{url}
\usepackage{graphicx}
\usepackage{setspace}
\usepackage{caption}
\usepackage{epstopdf}
\usepackage{cite}
\usepackage{amsmath}
\usepackage{amsthm}
\usepackage{float}
\usepackage{times}
\usepackage{enumerate}
\usepackage{amssymb}
\usepackage{algorithm}
\usepackage{algorithmic}
\usepackage{multirow}
\usepackage{braket}
\usepackage{multicol}
\usepackage{subcaption}

\newtheorem{definition}{Definition}
\newtheorem{remark}{Remark}

\IEEEoverridecommandlockouts

\begin{document}

\title{Network-Level Performance Evaluation of a Two-Relay
Cooperative Random Access Wireless System}

\author{\IEEEauthorblockN{Georgios Papadimitriou, Nikolaos
Pappas, Apostolos Traganitis, Vangelis Angelakis\thanks{
G. Papadimitriou is with Ericsson, Stockholm, Sweden (e-mail:
georgios.papadimitriou@ericsson.com).~N. Pappas and V. Angelakis
are with the Department of Science and Technology, Link\"{o}ping
University, Norrk\"{o}ping SE-60174, Sweden (e-mail:
\{nikolaos.pappas, vangelis.angelakis\}@liu.se).~A. Traganitis is
with the Computer Science Department, University of Crete, Greece
and the Institute of Computer Science, Foundation for Research and
Technology - Hellas (FORTH) (e-mail: tragani@ics.forth.gr).
This work was presented in part in IEEE ICC
2013~\cite{b:PapadimitriouICC2013}.
}
\thanks{The research leading to these results has received
funding from the People Programme (Marie Curie Actions) of the
European Union's Seventh Framework Programme FP7/2007-2013/ under
REA grant agreements n$^o$
[324515] (MESH-WISE) and n$^o$ [612361] (SOrBet).}
}}

\maketitle

\begin{abstract}
In wireless networks relay nodes can be used to assist the
users' transmissions to reach their destination. Work on relay
cooperation, from a physical layer perspective, has up to now
yielded well-known results. This paper takes a different stance
focusing on network-level cooperation. Extending previous
results for a single relay, we investigate here the benefits
from the deployment of a second one. We assume that the two
relays do not generate packets of their own and the system
employs random access to the medium; we further consider
slotted time and that the users have saturated queues. We
obtain analytical expressions for the arrival and service rates
of the queues of the two relays and the stability conditions.
We investigate a model of the system, in which the users are
divided into clusters, each being served by one relay, and show
its advantages in terms of aggregate and throughput per user.
We quantify the above, analytically for the case of the
collision channel and through simulations for the case of
Multi-Packet Reception (MPR), and we provide insight on when
the deployment of a second relay in the system can yield
significant advantages. 
\end{abstract}

\section{Introduction}
\label{sec:Into}

Cooperative communications have gained significant attention lately. Cooperation can take place in difference communication layers, with the bulk of interest focusing on physical layer performance \cite{b:Yates-NOW, b:Laneman}. In that level, cooperation benefits are self-evident, since the explored systems typically belong to a single actor with interest to maximize a specific utility \cite{b:NgYuJSAC2007}. Promoting cooperation at higher layers, has also drawn significant attention due to the potential benefits from operators and users. Focusing on the purely network layer the benefits of utilizing cooperative techniques have been recently shown to be multi-fold, with respect to system performance in terms of throughput \cite{b:Sadek, b:RongCooperation, b:Pappasitw, b:PappasTWC15, b:PappasISIT2012, b:PappasGlobalsip2013}, reliability \cite{b:LiuBOOK} and delay \cite{b:PappasTWC15}. In that regard the use of dedicated relays has been introduced in may practical systems, such as wi-fi (known as range extenders) and in LTE. 

\subsection{Related Work}

The notion of cooperative communications was introduced by information theory with the relay channel.
The relay channel is the basic building block for the
implementation of cooperative communications, which are widely
acknowledged to provide higher communication rates and
reliability in a wireless network with time varying
channels~\cite{b:Laneman}. It was initially proposed by van der
Meulen~\cite{b:vanderMeulen}, and its first information-theoretic
characterizations were presented in~\cite{b:CoverGamal}.

Recently, the study of the relay channel has gained significant
interest in the wireless communications community.
In~\cite{b:ZlatanovISIT} for the classic relay channel a protocol
is presented for selection of reception and transmission time
slots adaptively and based on the quality of the involved links.
Considering full-duplex and half-duplex
relaying~\cite{b:ZlatanovLETTER} shows
that if the numbers of antennas at source and destination are
equal to or larger than the number of antennas at the relay,
half-duplex relaying can achieve in some cases higher throughput
than ideal full-duplex relaying. With beamforming and taking
inter-relay interference ~\cite{b:SuMinKim}
proposes two buffer-aided relay selection
schemes. Interference cancellation is employed in~\cite{b:Nomikos} to allow
opportunistic relaying selection maximising the average capacity
of the network. For a practical system, OFDMA based cellular resource
allocation schemes are proposed in~\cite{b:PopovskiTVT08} for multiple 
relay stations (RS) with adaptive RS activation. 

As mentioned, the majority of the works in this area focus on potential gains by cooperation
on the physical layer. Recent works~\cite{b:Sadek}
and~\cite{b:RongCooperation} suggest that similar gains can be
achieved by network-layer cooperation. By network-layer
cooperation they consider relaying to be taking place at a
protocol level avoiding physical layer considerations.
Random multiple access schemes in these works use the collision channel model with erasures, 
where concurrent transmissions will fail \cite{b:RongCooperation, b:PappasGlobecom, b:PappasITW13}. 
The collision channel however is not the appropriate model for wireless networks.

Random access with Multi-Packet Reception (MPR) capabilities has
attracted attention recently \cite{b:Tong, b:Ghez, b:Naware, b:PappasCOMCOM}. The
seminal paper \cite{b:Tong} was the first to examine MPR as an
interaction between the physical and medium access control layers
for a wireless random access network. In \cite{b:Ghez}, the notion of MPR was introduced and two important
theorems for the slotted ALOHA network with MPR are provided.
They consider the effect of MPR on stability and delay of slotted
Aloha based random-access system and it is shown that the
stability region undergoes a phase transition from a concave
region to a convex polyhedral region as the MPR capability
improves in a two-user system. In \cite{b:Naware}, the
authors specify a general asymmetric MPR model and the medium
access control capacity region. In \cite{b:PappasCOMCOM}, the
impact of a relay node to a network with a finite number of
users-sources and a destination node is investigated. In this
network the relay and the destination nodes have MPR capabilities. Analytical expressions for the
characteristics of the relay node queue such as average queue
length, stability conditions etc. were obtained. Finally, an overview of
MPR-related research work covering the theoretically proved
impacts and advantages of using MPR from a channel perspective to
network capacity and throughput, the various technologies that
enable MPR from transmitter, transceiver, and receiver
perspectives and previous work on protocol improvement to better
exploit MPR, is provided in \cite{b:Lu}.

\subsection{Contribution}

In this work, we provide a thorough study of the impact of using
two relay nodes in a network to assist with relaying packets from
a number of users to a destination node. We first investigate the
system  analytically, assuming the collision channel; then we
move to assume that the system is MPR enabled and we conduct a
thorough, system-level simulation study. Our common assumptions
in both models are that (i) users have saturated queues and
random access to the medium with slotted time\footnote{Dealing with analytical performance evaluation of random access systems above three users with random arrivals is mathematically intractable. Specifically, assuming the sources saturated, the so-called saturated throughput can be obtained and is an inner bound of the stable throughput\cite{b:TonyUnion, b:LuoNqueues}.}; (ii) the
transmission of a packet takes the duration of exactly one time
slot; (iii) the two relays are dedicated, i.e. do not have
packets of their own, but assist the users by relaying their
packets when necessary; (iv) the wireless link between any two
nodes of the network is a Rayleigh narrowband flat-fading channel
with additive Gaussian noise.

In the first part, we obtain analytical expressions for the
arrival and service rates of the queues of the two relays, and
for the stability conditions. In doing so we use the stochastic
dominance technique~\cite{b:EphremidesStochastic} because the two
queues are coupled (i.e., the service process of each queue
depends on  the other queue having a packet to send or not). We
also look into a topology of the system in which the users are
divided into two clusters. In this scenario, we consider that the
users of one cluster do not interfere with the users or the relay
of the other cluster, still, the relays are interfering with each
other. This corresponds to the case of having the users in two
distant areas. However, since the location of the users is captured by the link success probability, 
this scenario can cover any similar case, in which a system practitioner could leverage sophisticated 
clustering techniques to approach our results, even in an on-line fashion. In general, clustering can 
deliver results depending on the topology of the users \cite{2014arXiv1403.4144H}.
For both scenarios (with and without clustering)
we study the impact of the two relay nodes of the two cases  on
the aggregate throughput and the throughput per user when the
queues of the two relays are stable. We show that the
probabilities of the two relays to attempt transmission do not
depend on each other when the queues are stable. The insertion of
the second relay offers a significant performance gain (higher
throughput) when the users are divided into clusters and each
cluster is assigned to one relay, though in the general
un-clustered scenario the gains are not as significant.

Under the  MPR model, the transmission of a node $j$ is
successful if the received Signal to Interference plus Noise
Ratio (SINR) is above a threshold $\gamma _j$. Here, due to queue
coupling the stability analysis and the derivation of analytical
expressions for the characteristics of the relays' queues such as
arrival and service rates, are not tractable. We therefore
conduct extensive simulations to provide a comprehensive insight
into the performance of the two-relay system. We show that the
use of two relays offers significant advantage in terms of
aggregate and throughput per user compared to systems with one
and no relay, for values of SINR threshold $\gamma>1$. Under the
clustering scenario employed in the first part we study the
impact on the aggregate and throughput per user compared to the
cases of no relay, one relay and two variations of two relay
nodes' operation:  a packet received by both relays is either
kept by (i) both nodes or (ii) by the one with the smallest queue
of the two. Finally, we provide insight for the average queue
size and the average delay per packet of the systems presented.

The paper is structured as follows: in Section \ref{sec:SystemModel} we describe the
system model. In Section \ref{sec:Analysis} we derive the analytical expressions
for: (A) the arrival and service rate of the relays' queues, (B)
the stability conditions of the queues, (C) the stability region
and (D) the throughput per user along with the upper and lower
bounds. In Section \ref{sec:4.CC-res} we present the numerical and simulation
study of the analytically obtained results of Section \ref{sec:Analysis}, while
in section \ref{sec:5.CC-res} we conduct a thorough simulation study for the MPR
model. We give our conclusions in Section \ref{sec:conclusions}.

\section{System Model}
\label{sec:SystemModel}
\subsection{Network Model}
\label{subsec:NetworkModel}
\begin{figure}[!t]
\centering
\includegraphics[scale=0.15]{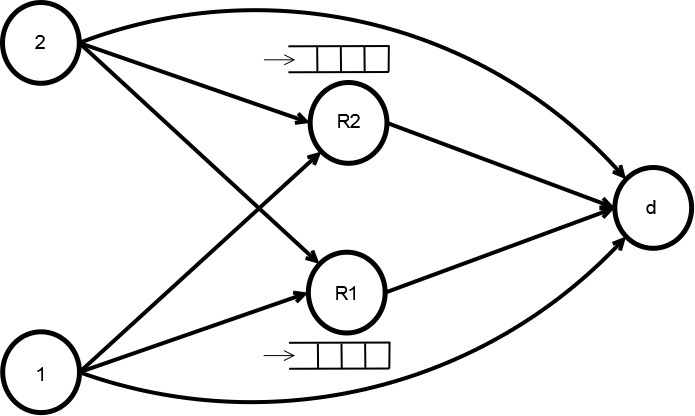}
\caption{Two relay nodes with $N=2$ user nodes.}
\label{fig1}
\end{figure}

We consider a network with $N$ source users, two relay nodes
$R_1$ and $R_2$ and a common destination node $d$, a case for
$N=2$ is depicted in Fig.~\ref{fig1}. The sources transmit
packets to the destination with the cooperation of the two
relays. We assume that the queues of the users are saturated.
The users have random access to the medium with no coordination
among them. The channel is slotted in time and the transmission
of a packet takes the duration of exactly one time slot. 
We assume fixed packet size, which could be viewed as an average packet size, 
since taking into account variable packet sizes would severely complicate the analysis.
The acknowledgements (ACKs) of successful transmissions are
instantaneous and error free. With this set of assumptions, and especially random access of the medium, 
a host of system parameters that could be available, such as channel state information for the links 
is not required nor considered in our work.

The relays do not generate packets of their own. If a
transmission of a user's packet to the destination fails, the
relays store it in their queues and try to forward it to the
destination at a subsequent time slot. In case that both relays
receive the same packet from a user, they choose randomly and
with equal probability which will store it in its queue. The queues
at the relays have infinite size.

In this work we consider two cases for the relays and the
destination, either that they are equipped with single
transceivers thus, a simultaneous transmission attempts by two or
more nodes (source-users or relays) result in a collision, or
that they are equipped with multiuser detectors, so that they may
decode packets successfully from more than one transmitter at a
time. The specifics for these are given in the following
section.

The notation we consider throughout this paper is the following:
The users attempt to transmit with probabilities $q_{i}$, where
$i=1, 2,\dots ,N$. Each of the relays, having not saturated
queues, attempts to transmit with probability $q_{R_j},~j=1,2$ if
its queue is not empty. Thus probability that a relay will
transmit a packet at a time slot $t$ is $q_{R_i}
\mathrm{Pr}(Q_{R_i}^{t} > 0)$, where $i = 1,2$ and $Q_{R_i}^{t}$
indicates the size of the queue at time slot $t$.

\subsection{Physical Layer Model}
\label{subsec:PHYModel}
\subsubsection{Collision Channel}
\label{sec:CCModel}
We model the link between two nodes $i$ and $j$ of the network as a
Rayleigh narrowband flat-fading channel with additive Gaussian
noise. The outage probability of that link with $SNR$ threshold
$\gamma_j$ is known~\cite{b:Tse} to be
$\mathrm{Pr}(SNR_{ij} < \gamma_j)=1-exp(- \gamma_j n_j
r_{ij}^{\alpha} / P_{tx}(i))$ where $P_{tx}(i)$ is the
transmission power of node $i$, $r_{ij}$ is the distance between
nodes $i$ and $j$, $\alpha$ is the path loss exponent and $n_j$
is the power of the additive white Gaussian noise at $j$. So, by
$p_{ij}$ we denote the success probability of a transmission
between nodes $i$ and $j$, which is $p_{ij}=exp(- \gamma_j n_j
r_{ij}^{\alpha} /P_{tx}(i))$.

The average service rate seen by the relay $R_1$ is
\begin{equation}\label{mR1}
\mu_{R_1}=q_{R_1}p_{R_1d} \left[1-q_{R_2} \mathrm{Pr}(Q_{R_2} >
0) \right] \prod_{i=1}^{N} (1-q_{i}).
\end{equation}
Because of the collision channel, all the users should remain
silent, which is with probability $\prod_{i=1}^{N} (1-q_{i})$,
also the relay $R_2$ should remain silent, with
probability $1-q_{R_2} \mathrm{Pr}(Q_{R_2} > 0)$. Furthermore,
relay $R_1$ has to be active, with probability $q_{R_1}$ and the
transmission to the destination successful
with probability $p_{R_1d}$.

Similarly, the average service rate seen by relay $R_2$ is

\begin{equation}\label{mR2}
\mu_{R_2}=q_{R_2}p_{R_2 d} \left[1-q_{R_1} \mathrm{Pr}(Q_{R_1} >
0) \right] \prod_{i=1}^{N} (1-q_{i}).
\end{equation}

Since the average service rate of each queue depends on the state
of the other, the problem of coupled queues arises.
Thus, we will apply the stochastic dominance approach to bypass
this difficulty.

\subsubsection{MPR}
\label{sec:MPRModel}
In the wireless environment, the collision channel is
restrictive, since we can not have more than one successful
transmissions simultaneously. Thus, we also consider the MPR
channel model, which is a generalized form of the packet erasure
model \cite{b:PappasCOMCOM}. In the MPR case, a node's
transmission is successful if the SINR is above a certain
threshold.
More specifically, if there exists a set of $T$ nodes
transmitting in the same time slot and ${P_{rx}}(i,j)$ is the
signal power received from node $i$ at node $j$ (when $i$
transmits), then the SINR$(i,j)$ determined by node $j$ is given
by
\begin{equation*}
\mbox{SINR}(i,j) = \frac{{{P_{rx}}(i,j)}}{{{n_j} +
\sum\nolimits_{k \in T\backslash \{ i\} } {{P_{rx}}(k,j)} }},
\end{equation*}
where $n_j$ is the receiver noise power at $j$.

We assume that a packet transmitted by $i$ is successfully
received by $j$ if and only if SINR$(i,j) \ge {\gamma _j}$, where
$\gamma _j$ is a threshold characteristic of node $j$. Moreover,
the wireless channel is subject to fading. Let ${P_{tx}}(i)$ be
the transmitting power of node $i$ and $r_{ij}$ be the distance
between $i$ and $j$. Then, the power received by $j$ when $i$
transmit is $P_{rx}(i,j) = A(i,j)g(i,j)$, where $A(i,j)$ is a
random variable representing channel fading and under Rayleigh
fading it is exponentially distributed \cite{b:Tse}. The receiver
power factor $g(i,j)$ is given by $g(i,j) =
{P_{tx}}(i){r_{ij}^{ - \alpha }}$, where $\alpha$ is the path
loss exponent with typical values between $2$ and $4$. The average success probability of a packet over link $ij$ when the transmitting nodes are in $T$ is given by~\cite{b:Tse}
\begin{equation*}
P_{i/T}^{j}=\exp\left(-\frac{\gamma_{j}\eta_{j}}{v(i,j)g(i,j)}\right) \prod_{k\in T\backslash \left\{i,j\right\}}{\left(1+\gamma_{j}\frac{v(k,j)g(k,j)}{v(i,j)g(i,j)}\right)}^{-1},
\end{equation*}

where $v(i,j)$ is the parameter of the Rayleigh random variable for fading.
%The analytical derivation for this success probability can be found in~\cite{b:Tse}.

\begin{remark}
In this work, the MPR case is considered only with simulations since the
analytical expressions even for the case of one relay are rather
complicated~\cite{b:PappasCOMCOM}. Additionally, with small values of the SINR threshold $\gamma$ is more likely to have more successful
simultaneous transmissions comparing to larger $\gamma$. More specifically, if $\gamma<1$ it is
possible for two or more users to transmit successfully at the
same time, comparing to $\gamma>1$ which that probability is almost
zero in the considered system setup.
\end{remark}

\subsection{Queue Stability}
We adopt the definition of queue stability used in~\cite{Szpankowski:stability}.

\begin{definition}
Denote by $Q_i^t$ the length of queue $i$ at the beginning of timeslot $t$. The queue is said to be \emph{stable} if
\begin{equation*}\label{eqn:definition_stability}
    \lim_{t \rightarrow \infty} {Pr}[Q_i^t < {x}] = F(x)  \text{ and } \lim_{ {x} \rightarrow \infty} F(x) = 1.
\end{equation*}

If $\lim_{x \rightarrow \infty}  \lim_{t \rightarrow \infty} \inf {Pr}[Q_i^t < {x}] = 1$, the queue is \emph{substable}. If a queue is stable, then it is also substable. If a queue is not substable, then we say it is unstable.
\end{definition}

Loynes' theorem~\cite{b:Loynes} states that if the arrival and service processes of a queue are strictly jointly stationary and the average arrival rate is less than the average service rate, then the queue is stable.

\section{Analysis}
\label{sec:Analysis}
In this section we will present the analysis for the collision
channel model. We obtain analytical equation for the arrival and
the service
rate of the two relays and also the stability region of the
system. Additionally, we obtain the throughput per user as well
as the aggregate throughput of the system.

In order to proceed further we need to calculate the average
arrival rates at the queues of the relays.

There is an arrival at the queue of relay $R_1$ if both relays
are silent, only one user transmits, and its transmission is
successfully received by $R_1$ but not by the destination. When
both relays receive the packet then the first will store it in
its queue with probability $\frac{1}{2}$ otherwise the second
relay will store it.

The probability that both relays are silent depends on the state
of the queues at the relays. Both relays are silent when their
queues are empty, which happens with probability
$\mathrm{Pr}(Q_{R_1}=0,Q_{R_2}=0)$, when the $i$ relay has a
non-empty queue but the queue at the $j \neq i$ relay is empty
then the probability that both relays are silent is
$(1-q_{R_i})\mathrm{Pr}(Q_{R_i}>0,Q_{R_j}=0)$. The probability
that both relays are silent when their queues are not empty is
$(1-q_{R_1})(1-q_{R_2})\mathrm{Pr}(Q_{R_1}>0,Q_{R_2}>0)$. The
average arrival rate at the first relay, $\lambda_{R_1}$, is

\begin{align}
\lambda_{R_1} = \left[
\mathrm{Pr}(Q_{R_1}=0,Q_{R_2}=0)+(1-q_{R_1})\mathrm{Pr}(Q_{R_1}>0,Q_{R_2}=0)
+ (1-q_{R_2})\mathrm{Pr}(Q_{R_1}=0,Q_{R_2}>0)+ \right. \notag \\
\left. +(1-q_{R_1})(1-q_{R_2})\mathrm{Pr}(Q_{R_1}>0,Q_{R_2}>0)
\right] \sum_{i=1}^{N} q_{i}p_{iR_1}(1-p_{id})
\left[(1-p_{iR_2})+\frac{1}{2}p_{iR_2}\right] \prod_{j=1,j \neq
i}^{N} (1-q_{j}),
\end{align}
which after simple manipulation becomes

\begin{equation}
\begin{aligned}
\lambda_{R_1} = \left[ 1-
q_{R_2}\mathrm{Pr}(Q_{R_2}>0)-q_{R_1}\mathrm{Pr}(Q_{R_1}>0) -
q_{R_1} q_{R_2}\mathrm{Pr}(Q_{R_1}>0,Q_{R_2}>0)\right] \times \\
\times \sum_{i=1}^{N} q_{i}p_{iR_1}(1-p_{id})
\left[(1-p_{iR_2})+\frac{1}{2}p_{iR_2}\right] \prod_{j=1,j \neq
i}^{N} (1-q_{j}).
\end{aligned}
\end{equation}

Symmetrically, we have that the arrival rate at the second relay,
$\lambda_{R_2}$, is
\begin{equation}
\begin{aligned}
\lambda_{R_2} = \left[ 1-
q_{R_1}\mathrm{Pr}(Q_{R_1}>0)-q_{R_2}\mathrm{Pr}(Q_{R_2}>0) -
q_{R_1} q_{R_2}\mathrm{Pr}(Q_{R_1}>0,Q_{R_2}>0)\right] \times \\
\times \sum_{i=1}^{N} q_{i}p_{iR_2}(1-p_{id})
\left[(1-p_{iR_1})+\frac{1}{2}p_{iR_1}\right] \prod_{j=1,j \neq
i}^{N} (1-q_{j}).
\end{aligned}
\end{equation}

With the previous expressions for $\lambda_{R_1}$ and
$\lambda_{R_2}$ we cannot proceed further, since each rate
depends on the joint probability density function of the queues.
This is a well known non-tractable problem, and in order to bypass this
difficulty we will deploy the stochastic dominance
technique~\cite{b:EphremidesStochastic} in order to decouple the
queues. The stochastic dominance technique was initially
developed to overcome the intractability arising in the analysis
of the inseparable multidimensional Markov chain for finite-user
buffered slotted ALOHA\footnote{The stochastic dominance
technique was introduced in \cite{b:EphremidesStochastic}
however, a brief introduction can be found in
\cite{b:Sidiropoulos}.}.

\subsection{Computation of Arrival and Service Rate}
\label{subsec:A.Rates}
The stochastic dominance approach implies the construction of two
hypothetical dominant systems. In the first system, say
$\mathcal{S}_1$, the relay $R_1$ reverts to the transmission of
``dummy packets" with the same probability, when its queue is
empty. All the other characteristics and assumptions of the
original system remain exactly the same. Similarly, in the second
system $\mathcal{S}_2$ the relay $R_2$ reverts to the
transmission of ``dummy packets" with the same probability, when
its queue is empty.

\subsubsection{\underline{Dominant system $\mathcal{S}_1$ --
Relay $R_1$ transmits ``dummy packets"}}
\label{sec:S1}
In the first dominant system, the relay $R_1$ transmits ``dummy
packets" when its queue is empty, thus
$\mathrm{Pr}(Q_{R_1}>0)=1$. The average arrival rates
$\lambda_{R_1}$ and $\lambda_{R_2 }$ are

\begin{eqnarray}\label{eq:lambdaR1S1}
\lambda_{R_1} =  (1- q_{R_1}) \left[ 1-
q_{R_2}\mathrm{Pr}(Q_{R_2}>0)\right]\sum_{i=1}^{N}
q_{i}p_{iR_1}(1-p_{id})
\left[(1-p_{iR_2})+\frac{1}{2}p_{iR_2}\right] \prod_{j=1,j \neq
i}^{N} (1-q_{j})
\end{eqnarray} 
\begin{eqnarray}\label{eq:lambdaR2S1}
\lambda_{R_2} = (1- q_{R_1}) \left[ 1-
q_{R_2}\mathrm{Pr}(Q_{R_2}>0)\right] \sum_{i=1}^{N}
q_{i}p_{iR_2}(1-p_{id})
\left[(1-p_{iR_1})+\frac{1}{2}p_{iR_1}\right] \prod_{j=1,j \neq
i}^{N} (1-q_{j}).
\end{eqnarray}

\begin{figure}[!t]
\centering
\includegraphics[scale=0.4]{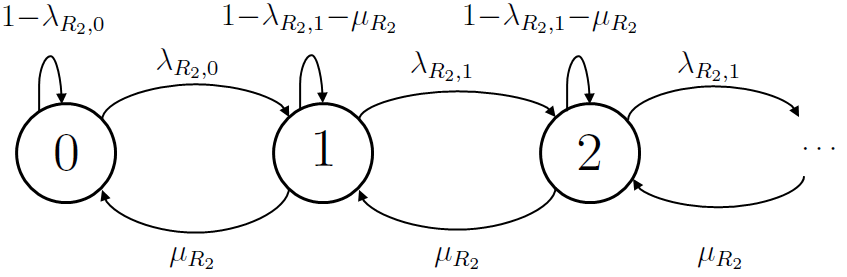}
\caption{The Discrete Time Markov Chain model of the queue at the relay $R_2$
for the first dominant system $\mathcal{S}_1$.}
\label{markov}
\end{figure}

The service rate of the relay $R_2$ in the system $S_1$ is given
by
\begin{equation} \label{eq:muR2S1}
\mu_{R_2}=q_{R_2}p_{R_2d}(1-q_{R_1}) \prod_{i=1}^{N} (1-q_{i}),
\end{equation}
and $\mu_{R_1}$ is given by (\ref{mR1}). The probability that the
queue at the $R_2$ is not empty is $\mathrm{Pr}(Q_{R_2}>0) =
\frac{\lambda_{R_2}}{\mu_{R_2}}$ which can be obtained from
Little's law. However, since the average arrival rate for the
second relay, $\lambda_{R_2}$, depends on the
$\mathrm{Pr}(Q_{R_2}>0)$ we cannot directly apply the previous
expression. Furthermore, $\lambda_{R_1}$ and the service rate
$\mu_{R_1}$ depend on the state of the queue of the second
relay.
Thus, we follow the procedure described
in~\cite{b:PappasGlobecom}. We model the queue at the $R_2$ as a
Discrete Time
Markov Chain (DTMC) with infinite states in order to describe the
queue evolution. The DTMC is depicted in Fig.~\ref{markov}.
The arrival rate of relay $R_2$ depends on whether its queue is
empty or not. If the queue is empty the arrival rate is denoted
by $\lambda_{R_2,0}$ and by $\lambda_{R_2,1}$ if it is not. Thus,
the average arrival rate $\lambda_{R_2}$ can be expressed also
as
\begin{equation}
\label{lambdaR2}
\lambda_{R_2}=\mathrm{Pr}(Q_{R_2}=0)\lambda_{R_2,0} +
\mathrm{Pr}(Q_{R_2}>0)\lambda_{R_2,1}.
\end{equation}
If the queue at the relay $R_2$ is empty then, we can easily show
that the probability of arrival $\lambda_{R_2,0}$ is
\begin{equation}
\label{lambdaR20}
\lambda_{R_2,0}=(1-q_{R_1}) \sum_{i=1}^{N}q_{i}p_{iR_2}(1-p_{id})
[(1-p_{iR_1})+\frac{1}{2}p_{iR_1}] \prod_{j=1,j \neq i}^{N}
(1-q_{j}).
\end{equation}
If the queue is not empty then the arrival rate is
\begin{equation}
\label{lambdaR21}
\lambda_{R_2,1}=(1-q_{R_2}) \lambda_{R_2,0}.
\end{equation}

By following the same methodology as in~\cite{b:PappasGlobecom}
and applying the method of balance equations, we can compute the
stationary distribution of the states $i$. The stationary
distribution of the Markov Chain exists if and only if
$\lambda_{R_2,1}<\mu_{R_2}$. Thus, the probability that the queue
at the $R_2$ is empty is given by
\begin{equation} \label{PQR2}
\mathrm{Pr}(Q_{R_2}=0)=\frac{\mu_{R_2} -
\lambda_{R_2,1}}{\mu_{R_2} - \lambda_{R_2,1} + \lambda_{R_2,0}}.
\end{equation}

From~(\ref{lambdaR20}),~(\ref{lambdaR21}),~(\ref{PQR2})
and~(\ref{lambdaR2}) we obtain
\begin{equation}
\label{lR2}
\lambda_{R_2}=\frac{\mu_{R_2}\lambda_{R_2,0}}{\mu_{R_2}-\lambda_{R_2,1}+\lambda_{R_2,0}}.
\end{equation}

Combining (\ref{lambdaR2}),~(\ref{lambdaR20}),~(\ref{lambdaR21})
and~(\ref{lR2}) we obtain the expression of the arrival rate
$\lambda_{R_2}$, shown in~(\ref{lR2final}), from which we see
that $\lambda_{R_2}$ does not depend on $q_{R_2}$, the
probability of transmission of $R_2$.
\begin{equation} \label{lR2final}
\lambda_{R_2}=\frac{p_{R_2d}\prod_{i=1}^{N}(1-q_{i})[\sum_{i=1}^{N}q_{i}p_{iR_2}(1-p_{id})(1-q_{R_1})\prod_{j=1,j
\neq
i}^{N}(1-q_{j})[(1-p_{iR_1})+\frac{1}{2}p_{iR_1})]}{p_{R_2d}\prod_{i=1}^{N}(1-q_{i})+\sum_{i=1}^{N}q_{i}p_{iR_2}(1-p_{id})\prod_{j=1,j
\neq i}^{N}(1-q_{j})[(1-p_{iR_1})+\frac{1}{2}p_{iR_1}]}.
\end{equation}

\subsubsection{\underline{Dominant system $\mathcal{S}_2$ --
Relay $R_2$ transmits ``dummy packets"}}
\label{sec:S2}
By following exactly the same procedure as in system
$\mathcal{S}_1$, we obtain the expressions for $\mu_{R_1}$,
$\lambda_{R_1}$, $\mu_{R_2}$ and $\lambda_{R_2}$.

\subsection{Necessary and Sufficient Stability Conditions}
\label{subsec:B.Stability}
The stability region of the system is defined as the set of
arrival rate vectors $(\lambda_{R_1},\lambda_{R_2})$ for which
the queues in the system are stable.
In order to derive the stability region we need to characterize
the average arrival rates $\lambda_{R_1}$ and $\lambda_{R_2}$ as
well as the average service rates $\mu_{R_1}$ and $\mu_{R_2}$. A
tool to obtain stability condition for a queue is the Loyne's
criterion \cite{b:Loynes}, which states that if the arrival rate
is less than the service rate then the queue is stable. The
average service rates are given by (\ref{mR1}) and (\ref{mR2})
thus the service rate of each queue depends on the state of the
other, thus we cannot apply the Loyne's criterion directly. We
will apply the stochastic dominance technique, which was
presented in the previous subsection for decoupling the queues
and to obtain necessary and sufficient conditions for the
stability. Recall that there are two dominant systems
$\mathcal{S}_1$ and $\mathcal{S}_2$. In $\mathcal{S}_i$ where in
the relay $R_i$ transmits dummy packets when its queue is empty
and all the other assumptions remain unaltered. Note that the
expressions for the average arrival and service rates change from
one dominant system to another since they depend on the
probability that a queue is empty.

In the first dominant system $S_1$, we have that the queues are
stable if $\lambda_{R_1}<\mu_{R_1}$ and
$\lambda_{R_2}<\mu_{R_2}$. The expression for
the service rate $\mu_{R_2}$ is given by (\ref{eq:muR2S1}) and
the service rate $\mu_{R_1}$ by (\ref{mR1}). Thus, using Little's
law we obtain that
\begin{equation} \label{eq:Q2EmptyS1}
\mathrm{Pr}(Q_{R_2}>0)=
\frac{\lambda_{R_2}}{q_{R_2}p_{R_2d}(1-q_{R_1}) \prod_{i=1}^{N}
(1-q_{i})}.
\end{equation}
After replacing (\ref{eq:Q2EmptyS1}) into (\ref{mR1}) we obtain
that
\begin{equation}
\mu_{R_1} = q_{R_1}p_{R_1d} \left(
1-q_{R_2}\frac{\lambda_{R_2}}{q_{R_2}p_{R_2d}(1-q_{R_1})
\prod_{i=1}^{N} (1-q_{i})} \right) \prod_{i=1}^{N}(1-q_{i}).
\end{equation}

Now we can apply Loyne's criterion for both queues and obtain the
region $\mathcal{R}_1$ from the first dominant system which is
given by
\begin{align} \label{eq:R1}
\mathcal{R}_1 = \left\{ (\lambda_{R_1},\lambda_{R_2}) :
\lambda_{R_1}<q_{R_1}p_{R_1d} \left(
1-q_{R_2}\frac{\lambda_{R_2}}{q_{R_2}p_{R_2d}(1-q_{R_1})
\prod_{i=1}^{N} (1-q_{i})} \right) \prod_{i=1}^{N}(1-q_{i}),
\right. \notag \\
\left. \lambda_{R_2}<q_{R_2}p_{R_2d}(1-q_{R_1})
\prod_{i=1}^{N}(1-q_{i}) \right\}.
\end{align}

From the above condition and after using (\ref{eq:lambdaR1S1}), (\ref{eq:lambdaR2S1}) and (\ref{lR2final}) we can further obtain the expression for
the transmission probability, $q_{R_1}$ where
\begin{equation}\label{qR1}
q_{R_1} > q_{R_1,min} \Leftrightarrow q_{R_1} >
\frac{\sum_{i=1}^{N}q_{i}p_{iR_1}(1-p_{id})\prod_{j=1,j \neq
i}^{N} (1-q_{j})\left[(1-p_{iR_2}) +
\frac{1}{2}p_{iR_2}\right]}{\sum_{i=1}^{N}q_{i}p_{iR_1}(1-p_{id})\prod_{j=1,j
\neq i}^{N} (1-q_{j})[(1-p_{iR_2})+\frac{1}{2}p_{iR_2}] +
p_{R_1d}\prod_{i=1}^{N}(1-q_{i})}.
\end{equation}
So, the $q_{R_1,min}$ for which the queue is stable is given by
\begin{equation}
\label{qR1min}
q_{R_1,min} =
\frac{\sum_{i=1}^{N}q_{i}p_{iR_1}(1-p_{id})\prod_{j=1,j \neq
i}^{N}
(1-q_{j})[(1-p_{iR_2})+\frac{1}{2}p_{iR_2}]}{\sum_{i=1}^{N}q_{i}p_{iR_1}(1-p_{id})\prod_{j=1,j
\neq i}^{N} (1-q_{j})[(1-p_{iR_2})+\frac{1}{2}p_{iR_2}] +
p_{R_1d}\prod_{i=1}^{N}(1-q_{i})}.
\end{equation}

From the second dominant system, $\mathcal{S}_2$, symmetrically
we obtain the stability region $\mathcal{R}_2$.
\begin{align} \label{eq:R2}
\mathcal{R}_2 = \left\{ (\lambda_{R_1},\lambda_{R_2}) :
\lambda_{R_2}<q_{R_2}p_{R_2d} \left(
1-q_{R_1}\frac{\lambda_{R_1}}{q_{R_1}p_{R_1d}(1-q_{R_2})
\prod_{i=1}^{N} (1-q_{i})} \right) \prod_{i=1}^{N}(1-q_{i}),
\right. \notag \\
\left. \lambda_{R_1}<q_{R_1}p_{R_1d}(1-q_{R_2})
\prod_{i=1}^{N}(1-q_{i}) \right\}.
\end{align}

Following exactly the same procedure as in $\mathcal{S}_1$ we
obtain expressions and bounds for $q_{R_2}$, and $q_{R_2,min}$
similar to~(\ref{qR1}) and~(\ref{qR1min}) respectively with $R_1$
and $R_2$ interchanged. Similarly to system $\mathcal{S}_1$, we
observe that $q_{R_2}$ does not depend on $q_{R_1}$. The queue of
the relay $R_2$ is stable if $q_{R_2}$ satisfies the inequality
\begin{equation}
q_{R_2,min} <q_{R_2}<1.
\end{equation}

\begin{figure}[!t]
\centering
\includegraphics[scale=0.5]{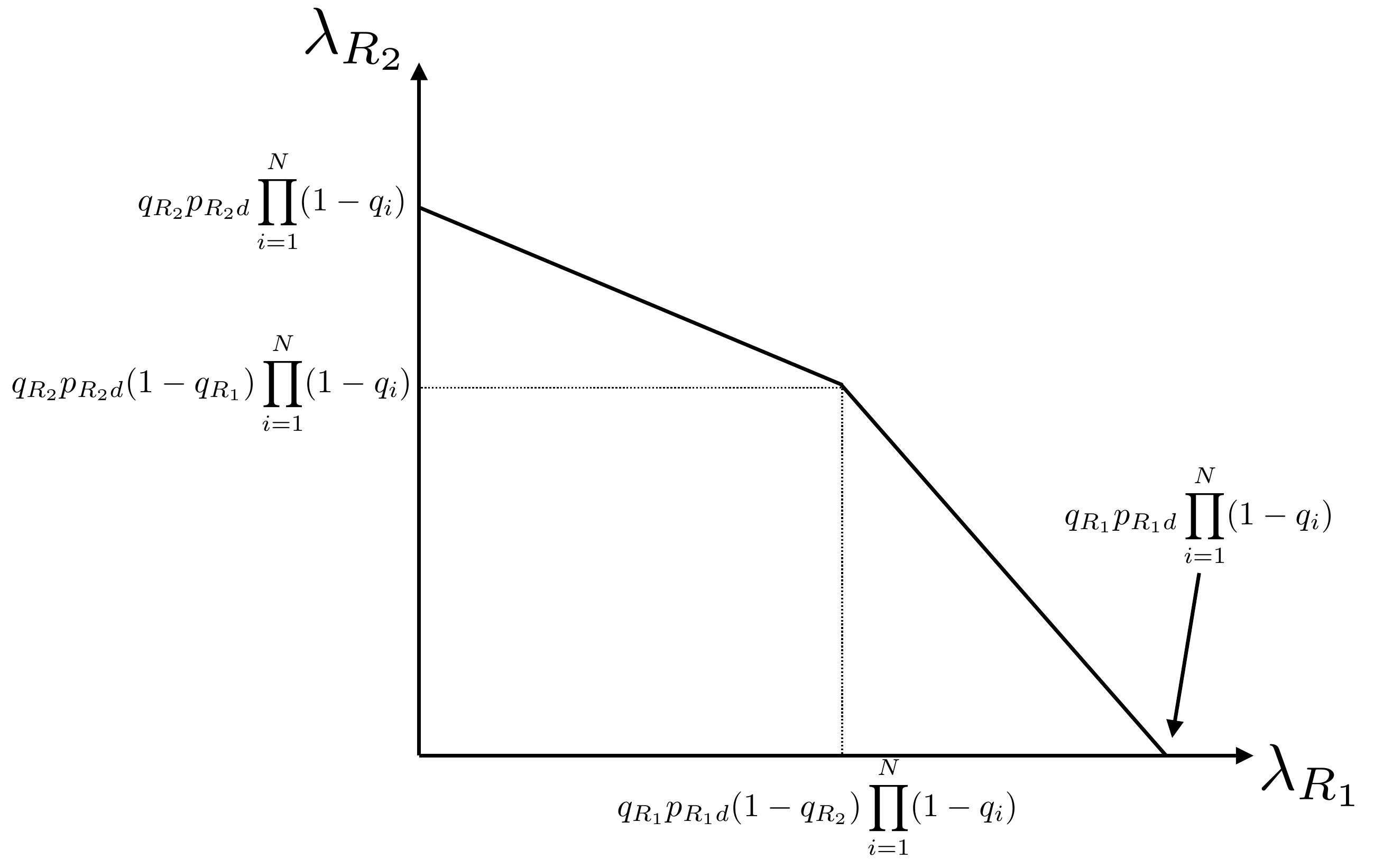}
\caption{The stability region, $\mathcal{R}$, of the system. The boundaries of
the region are given by (\ref{eq:R1}) and (\ref{eq:R2}).}
\label{stabregion}
\end{figure}

Finally the stability region of the system, $\mathcal{R}$, is
$\mathcal{R} = \mathcal{R}_1 \cup \mathcal{R}_2$ and is shown in
Fig.~\ref{stabregion}.

It is interesting to note that in \cite{b:EphremidesStochastic},
the stability conditions obtained by the dominant systems are not
merely sufficient, but sufficient and necessary for the stability
of the original system. The proof relies on the
\emph{indistinguishability} argument which also applies in our
case. By considering the properties of the dominant system $S_1$,
we can see that the queue sizes of the two relays cannot be
smaller than those in the original system, provided the queues
start with identical initial conditions in both systems. By
Loynes' Theorem, the stability condition of a queue is given by
$\lambda < \mu$.  Therefore, given that $\lambda_{R_2} <
\mu_{R_2}$, if for some $\lambda_{R_1}$ the queue $R_1$ in the
dominant system $S_1$ is stable, then the queue is also stable in
the original system. Conversely, if for some $\lambda_{R_1}$ in
the dominant system $S_1$ the queue $R_1$ is unstable, then it
will not transmit any ``dummy packets" and as long as the queue
does not empty, the dominant and the original systems behave
identically and as a consequence, the queue is unstable in the
original system as well.

\subsection{Throughput Per User}
\label{subsec:D.Throughput}
In this part, we will give the expression for the average user
throughput, which is the average rate of packets departing from
each user. There is a departure of a packet from a node if it
transmits whereas the two relays and all the other users are
silent, and its transmission is either successfully received by
the destination or if unsuccessful, it is successfully received
by $R_1$ or $R_2$. Thus, the throughput rate $\mu_{i}$ seen by
the user $i$ is
\begin{align}
\mu_{i}=q_{i}\left[p_{id}+(1-p_{id})(p_{iR_1}+p_{iR_2}-p_{iR_1}p_{iR_2})\right]
\left[
\mathrm{Pr}(Q_{R_1}=0,Q_{R_2}=0)+(1-q_{R_1})\mathrm{Pr}(Q_{R_1}>0,Q_{R_2}=0)+
\right. \notag \\
\left. + (1-q_{R_2})\mathrm{Pr}(Q_{R_1}=0,Q_{R_2}>0)+
(1-q_{R_1})(1-q_{R_2})\mathrm{Pr}(Q_{R_1}>0,Q_{R_2}>0) \right]
\prod_{j=1,j \neq i}^{N}(1-q_{j}),
\end{align}
which, after some simplifications is given by
\begin{equation}
\label{eq:thrperuser}
\begin{aligned}
\mu_{i}=q_{i}
\left[p_{id}+(1-p_{id})(p_{iR_1}+p_{iR_2}-p_{iR_1}p_{iR_2})\right]
\times \\
\times \left[ 1-
q_{R_2}\mathrm{Pr}(Q_{R_2}>0)-q_{R_1}\mathrm{Pr}(Q_{R_1}>0) -
q_{R_1}
q_{R_2}\mathrm{Pr}(Q_{R_1}>0,Q_{R_2}>0)\right]\prod_{j=1,j \neq
i}^{N}(1-q_{j}).
\end{aligned}
\end{equation}

We observe that the throughput per user depends on whether both
queues are empty or not. So, it is not tractable to find an
explicit expression of the throughput per user in closed form.
Instead, we will find an upper and a lower bound and by
simulation we will study the tightness of these bounds.

In order to find an upper bound, we will consider the case when
the two relays do not interfere with the users. This provides an
upper bound because if the relays do not interfere with the
users, the interference in the system is less and thus we get
higher throughput per user. This is the case when we assume that
the relays operate in a different channel than the users. This
upper bound is given by
\begin{equation}
\label{eq:simplethrperuserupper}
\mu_{i,upper}=q_{i}
\left[p_{id}+(1-p_{id})(p_{iR_1}+p_{iR_2}-p_{iR_1}p_{iR_2})\right]
\prod_{j=1,j \neq i}^{N} (1-q_{j}).
\end{equation}

In order to find a lower bound, we will assume that the two
relays have always packets in their queues, their queues never
empty. This can be the case when the relays are highly utilized.
The lower bound is given by
\begin{equation}
\label{eq:simplethrperuserlower}
\mu_{i,lower}=\mu_{i,upper}(1-q_{R_1})(1-q_{R_2}).
\end{equation}

However, the way we treated the relays so far, that can be
reached from any user, can be sub-optimal thus, in the following
subsection we consider the case of using one relay per cluster of
users.

\subsection{Improving the Throughput Per User by Clustering
Users}
\label{subsec:E.Clusters}
In order to improve the throughput per user of the system, we
consider the case that we divide the users into two clusters
served by relays $R_1$ and $R_2$. We assume that due to the
distance between clusters the users of the first cluster do not
interfere with the users of the second cluster at their relay. If
two users transmit simultaneously we will have a collision at the
destination. We also assume that when a relay transmits
simultaneously with the users, the users' transmissions do not
affect the relay's transmission to the destination node whereas
their transmissions to the destination fail. That is because of
the shorter distance between the relay and the destination and
also the higher transmit power of the relay compared to that of
the users'. Furthermore, when both relays transmit simultaneously
we have a collision at the destination. We divide the users
equally to both clusters and we assume that each cluster has
$N_{k}$ users with $k=1,2$ where $N_{1}=N_{2}=\frac{N}{2}$.

The throughput per user of the system described depends again on
whether both queues are empty or not. Thus, we find an upper and
a lower bound and we will show that the results of the simulation
of that system lie between those two bounds.
The upper bound of the throughput per user $i$ of cluster $k$ is
given by:

\begin{equation}
\label{clusterupper}
\mu_{i,k,upper}=q_{i}p_{id} \prod_{j=1,j \neq i}^{N} (1-q_{j}) +
q_{i}(1-p_{id})p_{iR_k} \prod_{j=1}^{N_{k}} (1-q_{j}).
\end{equation}

The lower bound for the throughput is given by
\begin{equation}
\label{clusterlower}
\mu_{i,k,lower}=\mu_{i,k,upper}(1-q_{R_1})(1-q_{R_2}).
\end{equation}

\begin{remark}
As presented earlier, the throughput per user is given by (\ref{eq:thrperuser}). In order to obtain the inner bounds given by (\ref{eq:simplethrperuserlower}) and (\ref{clusterlower}) we did the assumption that the relays have
saturated queues. These bounds become tight when the relays' queues approach saturation. The outer bounds for the throughput can be obtained by assuming that 
the relays' queues are always empty thus, the relays do not cause interference to the users' transmissions. Apparently, the obtained outer bounds are tight
when the queues at the relays are underutilized.
\end{remark}

%%%%%%%%%%%%%%%%%%%%%%%%%%%%%%%%%%%%%%%%%%%%%%%%%%%%%%%%%%%%%%%%%%%%%%%%%%%%%%%%%%%%%%%%%%%%%%%%%%%%%%%%%

%%%%%%%%%%%%%%%%%%%%%%%%%%%%%%%%%%%%%%%%%%%%%%%%%%%%%%%%%%%%%%%%%%%%%%%%%%%%%%%%%%%%%%%%%%%%%%%%%%%%%%%%%
\section{Numerical and Simulation Results for the Collision
Channel Model}
\label{sec:4.CC-res}
In this section, we present the numerical results for the per
user and aggregate throughput of the system with two stable
relays for the collision channel model. We directly verify (in
Fig.~\ref{fig:PUThrpt}) that the throughput per user for the
cases of two relays lies between the upper and lower bounds given
in (22)-(25). Then, we compare these two cases with the system
without relay and the system with one relay.
The results presented below are averages of at least 10,000 runs
on each scenario verifying the accuracy of the analysis in the
previous sections. We consider that all $N$ users and both the
relays have the same link characteristics and transmission
probabilities for both the simple and the scenario with user
clustering. All parameters in our testing are given in
Table~\ref{tab:table1}.

\begin{figure}[!t]
\centering
\includegraphics[scale=0.5]{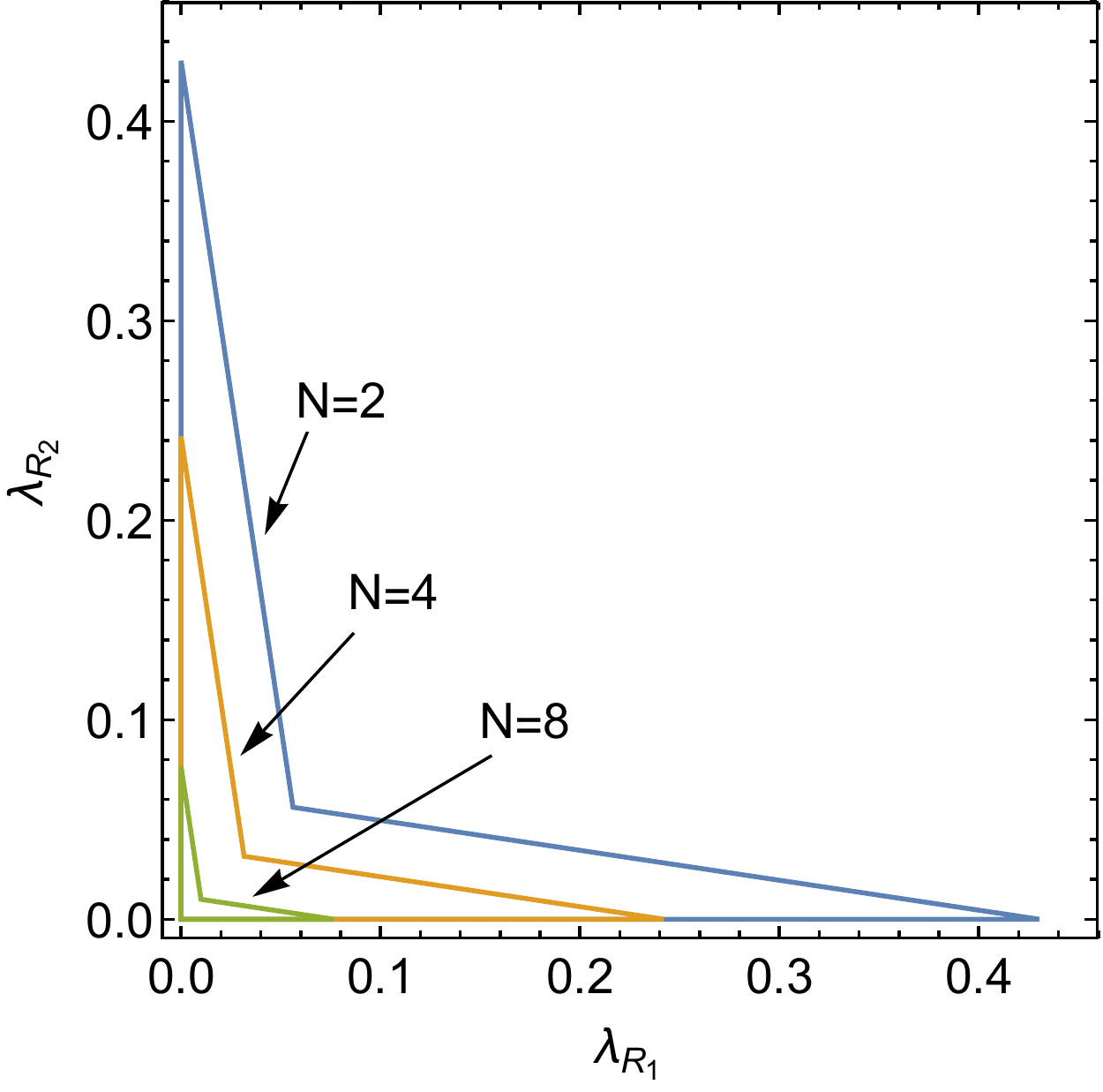}
\caption{The stability region of the system described in Table \ref{tab:table1} for $N=2,4,8$
users.}
\label{fig:stabregionplot}
\end{figure}

The stability region for the considered scenario for $N=2,4,8$
users is depicted in Fig. \ref{fig:stabregionplot}.
As the number of users increases we see that the boundary of the
region shrinks, which is expected since the number of collisions
increases in the network.

\subsection{Throughput Per User}
\label{subsec:PUthrp}
\begin{figure}[!t]
\centering
\begin{subfigure}[b]{0.55\textwidth}
\includegraphics[width=\textwidth]{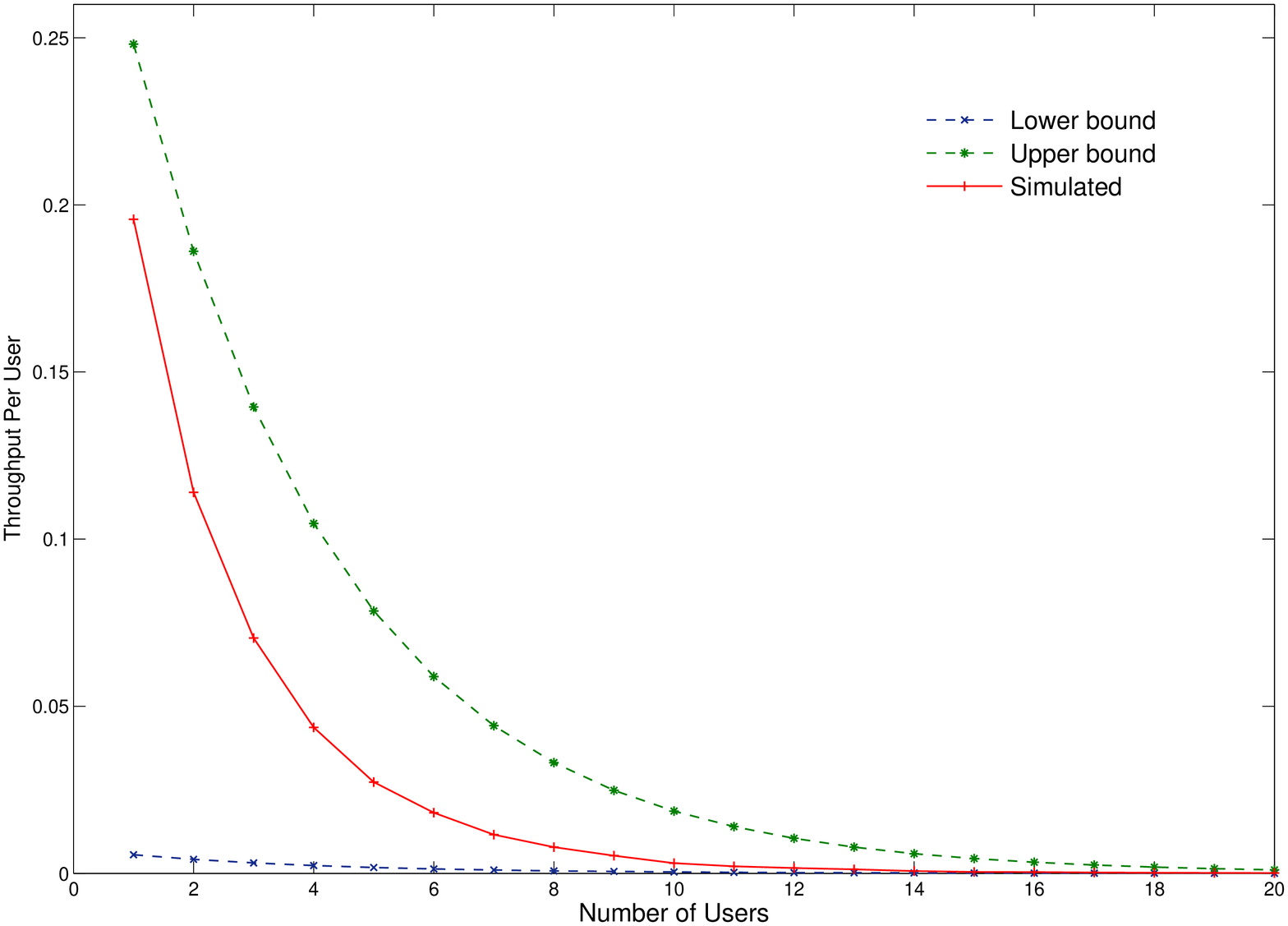}
\caption{``Simple'' Scenario}\label{subfig:SimpleThrpt}
\end{subfigure}
\begin{subfigure}[b]{0.55\textwidth}
\includegraphics[width=\textwidth]{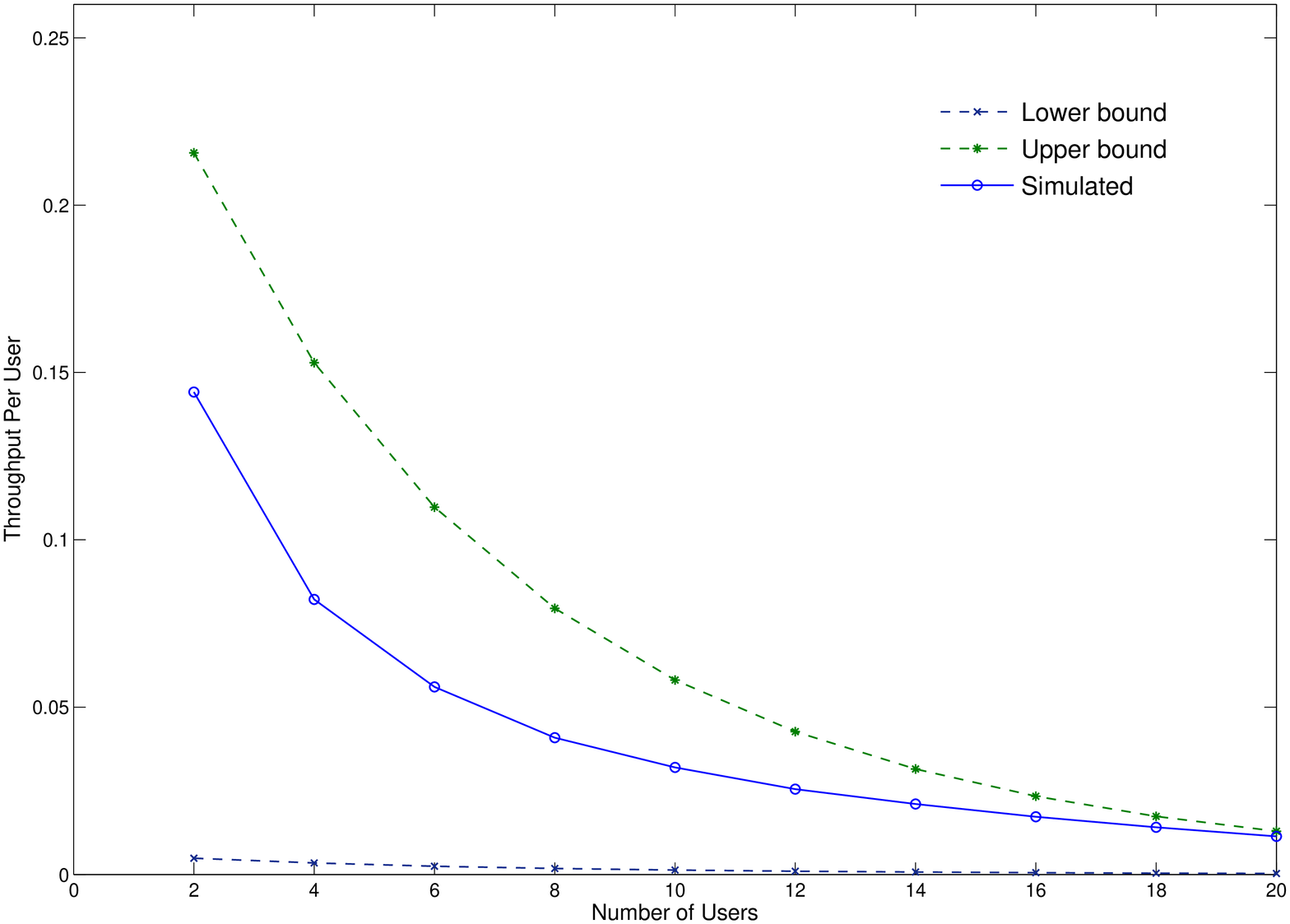}
\caption{``Clustering'' Scenario}\label{subfig:ClustersThrpt}
\end{subfigure}
\caption{Throughput per user vs. the number of users.}
\label{fig:PUThrpt}
\end{figure}

The plots of Fig.~\ref{fig:PUThrpt} present the throughput per
user versus the number of users of the ``simple'' scenario
described in Section~\ref{subsec:D.Throughput} and the
``clustering'' scenario of Section~\ref{subsec:E.Clusters}. As
expected the simulations lie between the lower and the upper
bounds defined in previous sections. In
Fig.~\ref{subfig:SimpleThrpt}, when the number of the users in
the system increases, we see that the throughput per user tends
to the lower bound. This is because the relays' queues are approaching
saturation as the number of users increases. In this case the lower bound becomes tight.

In Fig.~\ref{subfig:ClustersThrpt}, the
throughput per user tends to the upper bound, because the relays' queues tend to be empty most of the time. 
Thus, we have a better utilization of the system with clustering due to the
reduction of concurrent transmissions per relay resulting in less
collisions. Furthermore, one can observe in the simulation
curves, the result of orthogonalizing user transmissions to the
relays via clustering as an effective doubling of the throughput
for more than $4$ users.

\subsection{The Benefit of Using a Second Relay}
\label{subsec:2ndRelayCC}
\begin{figure}[!t]
\centering

\begin{subfigure}[b]{0.55\textwidth}
\includegraphics[width=\textwidth]{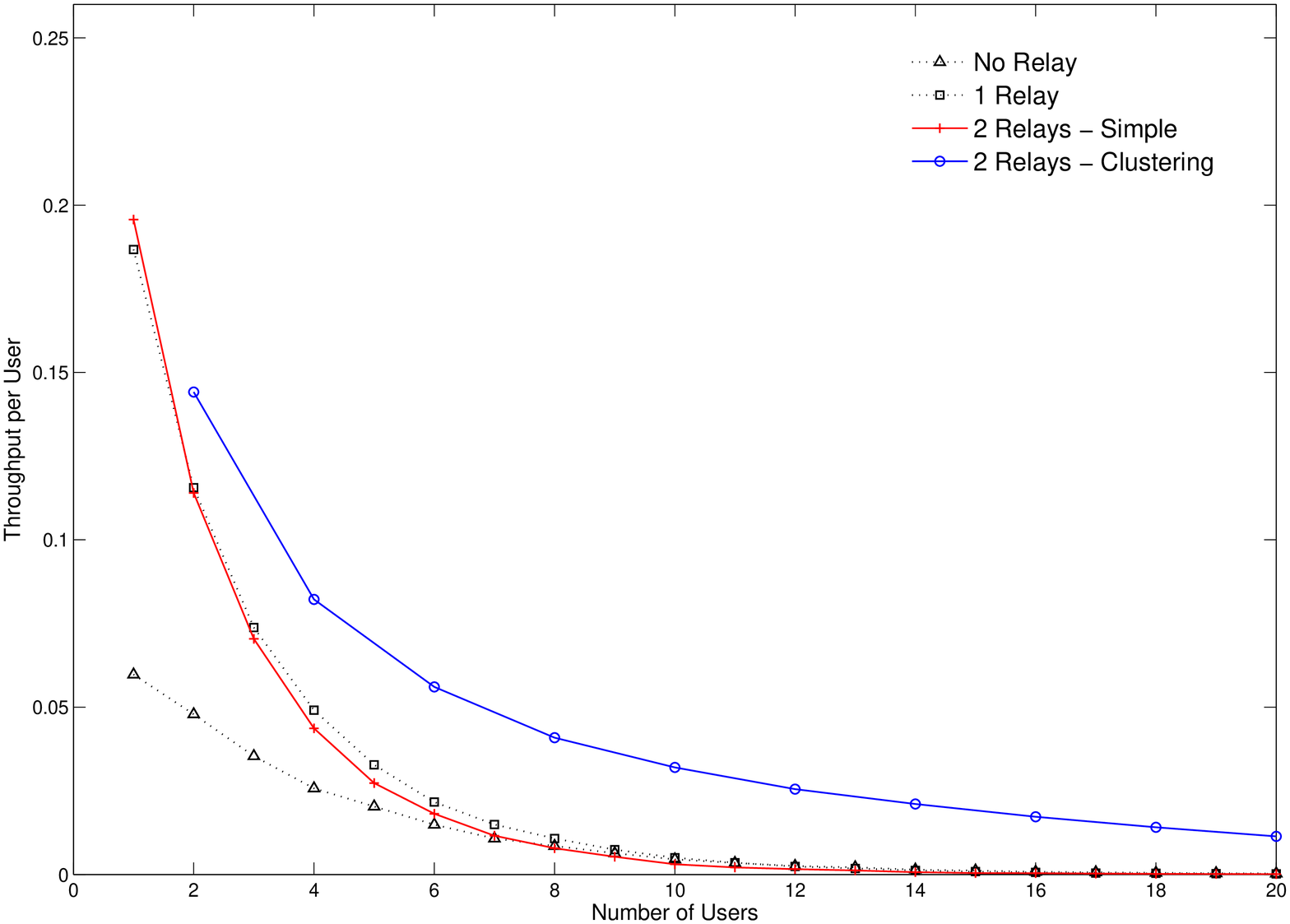}
\caption{Throughput per user}\label{subfig:AllThrptSim}
\end{subfigure}
\begin{subfigure}[b]{0.55\textwidth}
\includegraphics[width=\textwidth]{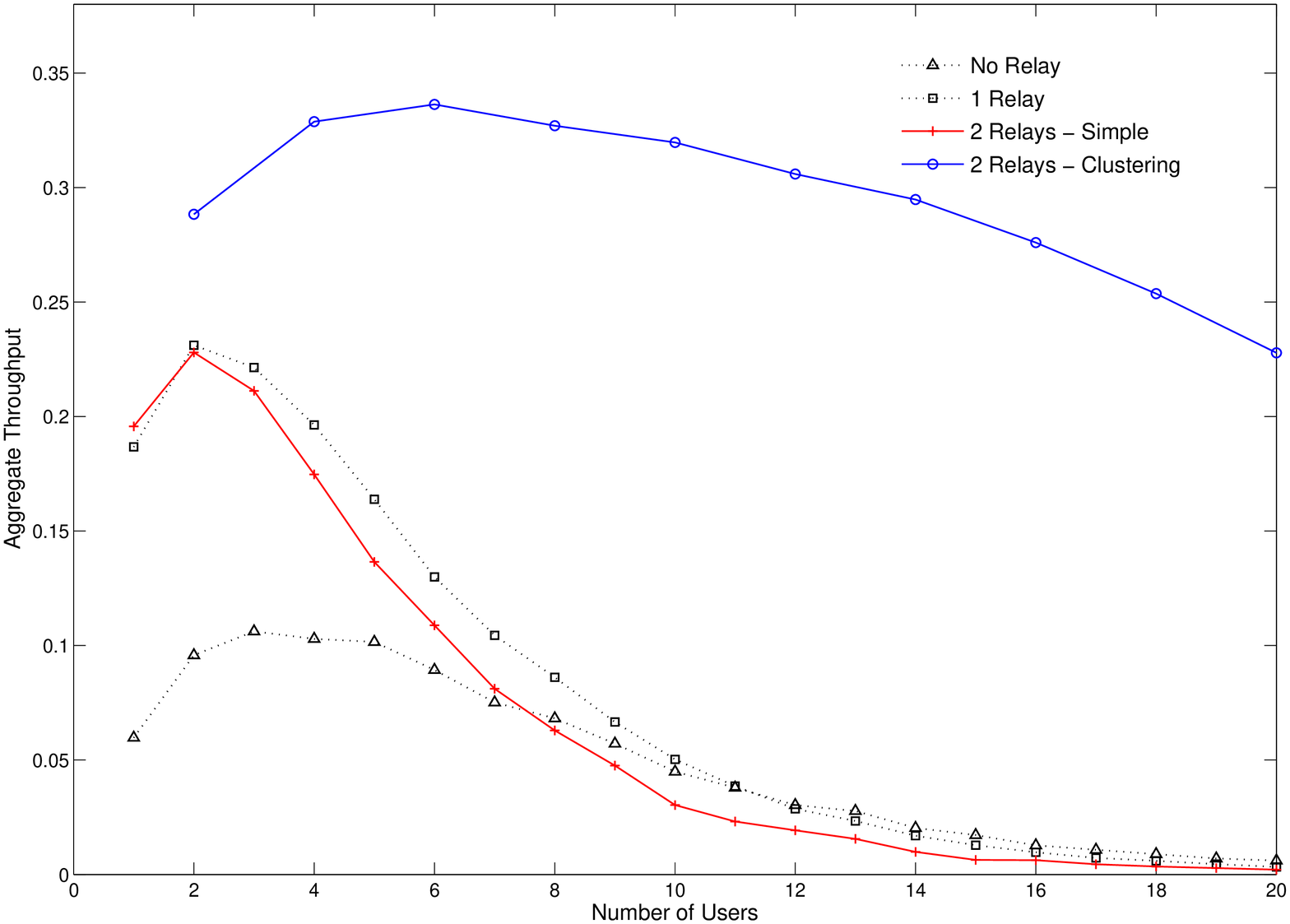}
\caption{Aggregate throughput}\label{subfig:SumThrpt}
\end{subfigure}
\caption{Comparisons of throughput vs. number of users, with the
Simple and Clustering Scenarios, against one relay and no
relay.}\label{fig:CompThrpt}
\end{figure}

The plots of Fig. \ref{fig:CompThrpt} present the per user and
aggregate throughput versus the number of users for the cases of
no relay, one relay and two relays (with and without clustering),
obtained by simulation. We observe that the simple system with
two relays does not offer any advantage over the system with one
relay. This is expected because the insertion of a second relay
with high probability to attempt transmission, when its queue is
not empty, generates more interference in the system. However,
this interference is alleviated in the clustering scenario which
thus offers significant advantage over the system with one relay
(more than 300\% higher aggregate throughput in our specific
setup).

\begin{table}[t!]
\centering
\caption{Parameters for the collision channel model
results.}%\vspace{-2mm}
\label{tab:table1}
\centering
\begin{tabular}{ l|p{5cm}|p{3.5cm}|p{3.5cm}}

\hline
\hline
\em{Notation}&
\em{Explanation} &
\em{Value in ``Simple'' Scenario}&
\em{Value in ``Clustering'' Scenario}\\

\hline
\hline

$p_{iR_j},~i=1,\dots N,~j=1,2$&
Success probability of transmission from user $i$ to relay $j$&
$p_{iR_1}=p_{iR_2}=0.9,~i=1\dots  N$&
$p_{iR_1}=p_{jR_2}=0.9,~i=1\dots N_1,~j=1\dots N_2$\\

\hline
$p_{R_jd},~j=1,2$ &
Success probability of transmission from relay $j$ to the
destination &
\multicolumn{2}{c}{$p_{R_jd}=0.9,~j=1,2$}\\
% \multicolumn{2}{c}{$p_{iR_1}=p_{iR_2}=0.9$}

\hline
$p_{id},~i=1,\dots N$  &
Success probability of transmission from user $i$ to the
destination &
\multicolumn{2}{c}{$p_{id}=0.25$} \\

\hline
$q_{R_j},~j=1,2$&
Probability that relay attempts to transmit in a timeslot, (if
its queue is not empty)&
\multicolumn{2}{c}{$q_{R_1}=q_{R_2}=0.85$}\\

\hline
$q_{i},~i=1,\dots  ,N$ &
Probability that user $i$ attempts to transmit in a timeslot &
\multicolumn{2}{c}{$q_{i}=0.25,i=1,\dots ,N$}\\

\hline
\hline
\end{tabular}
\end{table}

\begin{figure}[!t]
\centering
\includegraphics[scale=0.65]{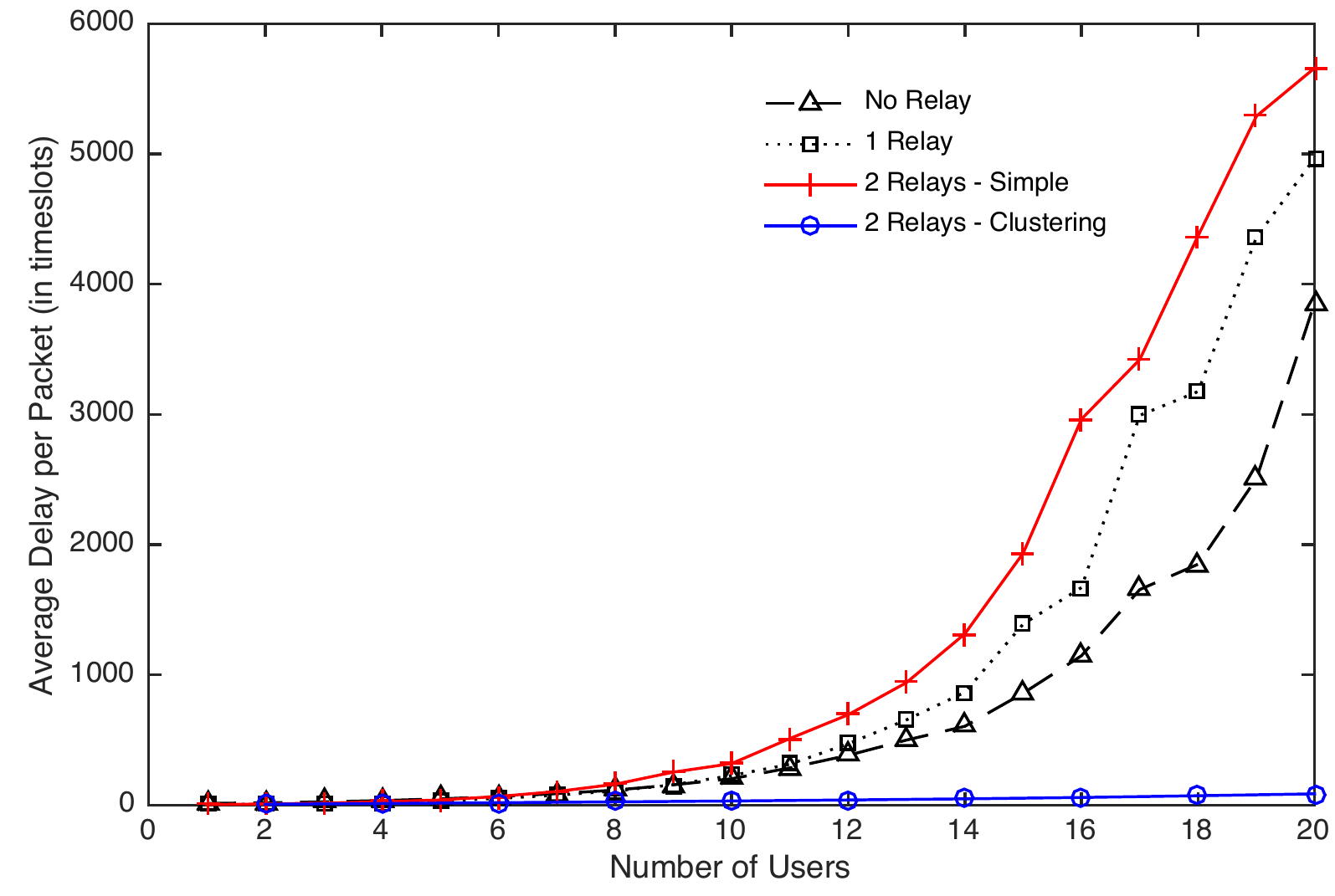}
\caption{Comparison of average per packet delay (in timeslots) vs. number of users, with Simple 
and Clustering scenarios, against one relay and no relay.}
\label{fig:CompDelayCollision}
\end{figure}

\subsection{Average Per Packet Delay}
Another important parameter in cooperative systems is the average
delay per packet. By delay we refer to the time it takes from the
moment a packet has been transmitted until it is delivered to the
destination. This parameter is important especially in
delay-sensitive networks for real-time services. Thus, here we
investigate the average delay per packet of systems with two
relays and compare it with the systems with one and no relay.

Fig. \ref{fig:CompDelayCollision} presents the average delay per packet (counted in timeslots) versus the number of users for the cases of
no relay, one relay and two relays (with and without clustering), obtained by simulation.

The clustered system with two relays provides the lowest average delay compared to the 
other systems. Furthermore, the clustered systems appears to be more prone to the increase
of the number of the users. Above eight users, the no relay system faces lower average delay than the one and the simple two-relay systems. The reason is that the system without relay does not suffer from additional delays introduced by packets queueing at the relay; this queueing delay increases with the number of users affecting the average per packet delay.

\section{Simulation Results for the MPR model}
\label{sec:5.CC-res}

\subsection{The Performance Benefits of Using a Second Relay}
\label{subsec:2ndRelayMPR}
Here we present the aggregate and throughput per user for the
cases with no relay, one relay, and two relays in the system for
the system with MPR, under different values of an assumed SINR
threshold $\gamma$. We examine two strategies to handle a user's
packet successfully received at both relays: either that
\textbf{(a)} both relays will store and forward it to the
destination ({\em Simple}) or that \textbf{(b)} the packet is
stored by the relay which has the smaller queue size ({\em
Smaller Queue Stores Packet}). If the queue size of the two
relays is equal, then the two relays choose randomly and with
equal probability which one will store the packet in its queue.
Furthermore, as we previously did for the collision channel model
we also study the potential impact of dividing the users into two
clusters served by relays $R_1$ and $R_2$.

\begin{figure}[!t]
\centering
\includegraphics[scale=0.25]{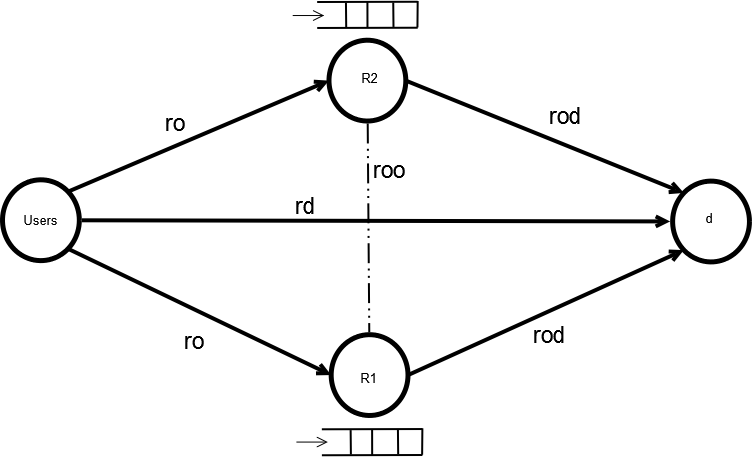}
\caption{Two relay nodes with $N$ users with same link
characteristics and transmission probabilities.}
\label{fig:2relays}
\end{figure}

An example topology of a two-relay test network with $N$
collocated users is depicted in Fig.~ \ref{fig:2relays}. The
parameters used in the simulations for each of the three cases
are shown in Table ~\ref{tab:table2}. To simplify the
presentation, we consider that all users have the same
transmission probabilities and all links to have the same SINR
threshold $\gamma$. Note that, with small values of $\gamma$ it
is more likely to have more successful simultaneous transmissions
comparing to larger $\gamma$. For $\gamma<1$ the probability for
two or more nodes to transmit successfully at the same time is
higher than the same probability when $\gamma>1$, which tends to
zero~\cite{b:PappasCOMCOM}.

\begin{table}[t!]
\centering
\caption{Simulation parameters for the MPR model results.}
\label{tab:table2}
\centering
\begin{tabular}{p{3cm}|p{5cm}|p{3.5cm}}
\hline
\hline
\em{Notation}&
\em{Explanation} &
\em{Value}\\

\hline
$r_d$&
users-destination distance&
100m\\

\hline
$r_R$&
users-relays distance&
59m\\

\hline
$r_{R'}$&
clustered user-non-serving relay distance&
88m\\

\hline
$r_{0d}$&
relays-destination distance&
59m\\

\hline
$r_{00}$&
inter-relay distance&
60m \\

\hline
$\alpha_{id}$&
users-destination path loss exponent&
$4$ \\

\hline
$\alpha_{i0}$&
users-relays path loss exponent&
$2$ \\

\hline
$\alpha_{0d}$&
relays-destination path loss exponent&
$2$ \\

\hline
$P_{tx}(i),~i=1\dots N$&
Transmit power of user $i$&
1mW\\

\hline
$P_{tx}(R_j),~j=1,2$&
Transmit power of each relay&
5mW\\

\hline
$q_i,~i=1\dots N$&
 Probability that user $i$ attempts to transmit in a timeslot &
$q_i=0.25,~i=1\dots N$\\

\hline
$q_{R_j},~j=1,2$&
Probability that relay $j$ attempts to transmit in each timeslot
(if its queue is not empty)&
$q_{R_1}=q_{R_2}=0.85$ \\

\hline
\hline
\end{tabular}
\end{table}

We note the assumptions made in our simulations. First, that the
path loss exponent between users-destination as well as between
the two relays is 4 while between users-relays and
relays-destination is taken to be 2. This so that the relay nodes
are more accessible for the users than the destination node. Thus
we consider user-relay and user-destination channels that are
more reliable than the user-destination one. Otherwise, the
presence of the relays would degrade the performance of the
network \cite{b:PappasCOMCOM}. We also assume that the transmit
power of the relays is five times higher than that of the users.
For the ``Smaller Queue Stores Packet'' strategy, we assume that
the relays communicate in a separate channel and thus these
transmissions do not interfere with those of the system we study.
For the clustering scenario, we divide the users equally to both
clusters and assume that relay $R_1$ cannot receive packets from
users of cluster 2 and relay $R_2$ cannot receive packets from
users of cluster 1 respectively, this is achieved by first taking
the respective path loss exponents to be equal to 4 and the
distance between cluster $1$ and relay $R_2$ to be $1.5$ times
the distance between cluster $1$ and relay $R_1$, and
vice-versa.

\begin{figure}[!t]
\centering
\begin{subfigure}[b]{0.45\textwidth}
\includegraphics[width=\textwidth]{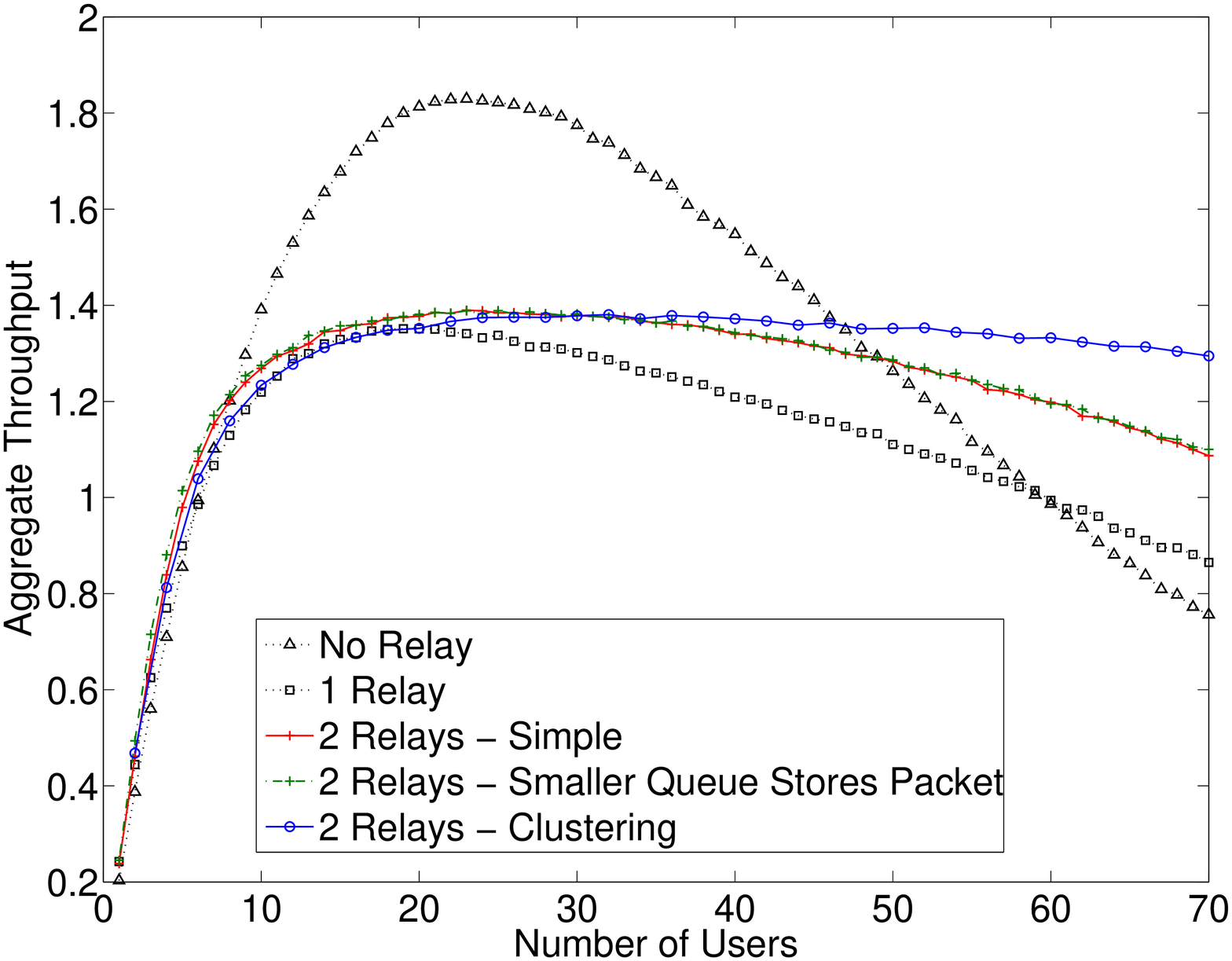}
\caption{$\gamma = 0.2$}\label{subfig:MPRg0.2agg}
\end{subfigure}
\begin{subfigure}[b]{0.45\textwidth}
\includegraphics[width=\textwidth]{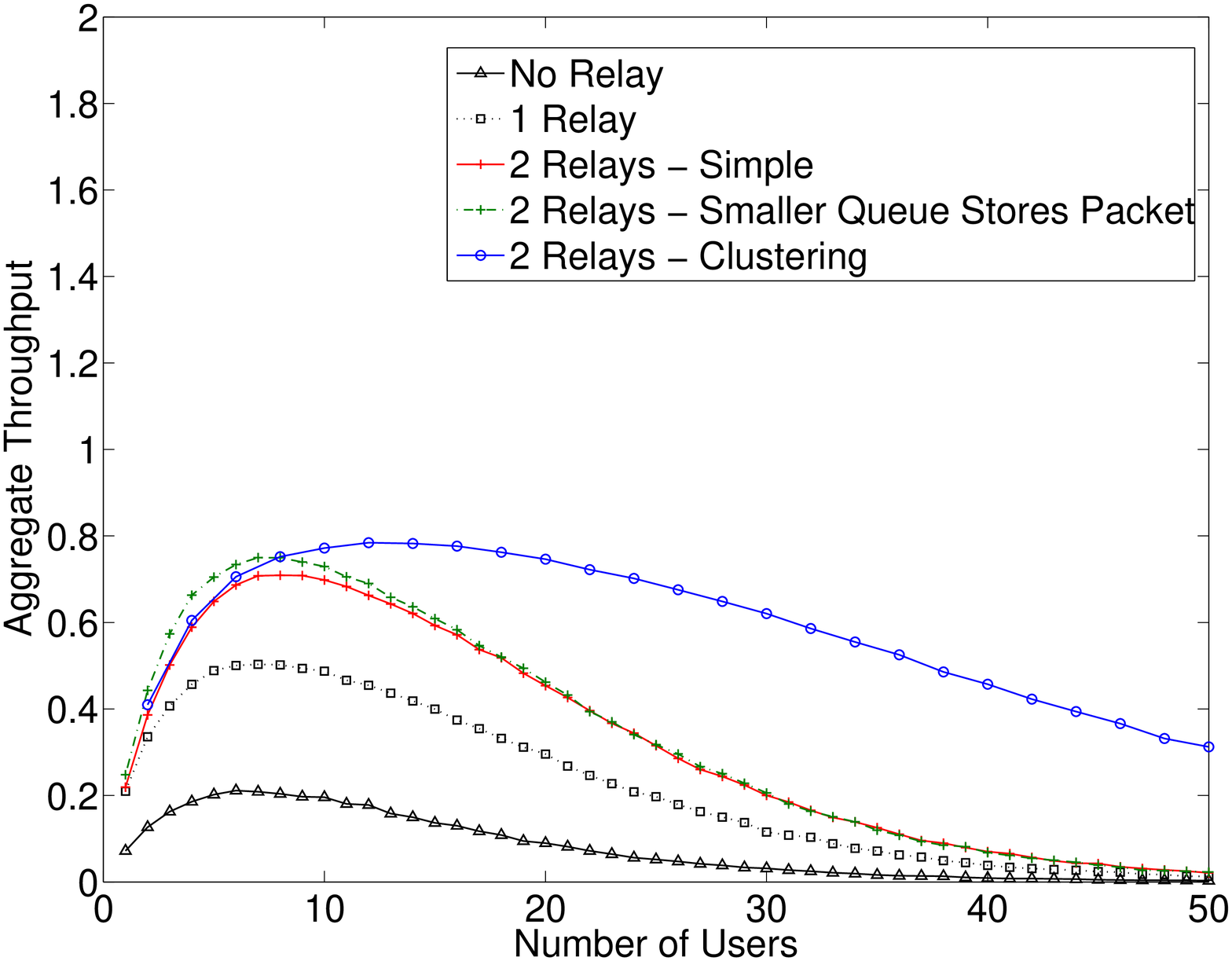}
\caption{$\gamma = 1.2$}\label{subfig:MPRg1.2agg}
\end{subfigure}
\begin{subfigure}[b]{0.45\textwidth}
\includegraphics[width=\textwidth]{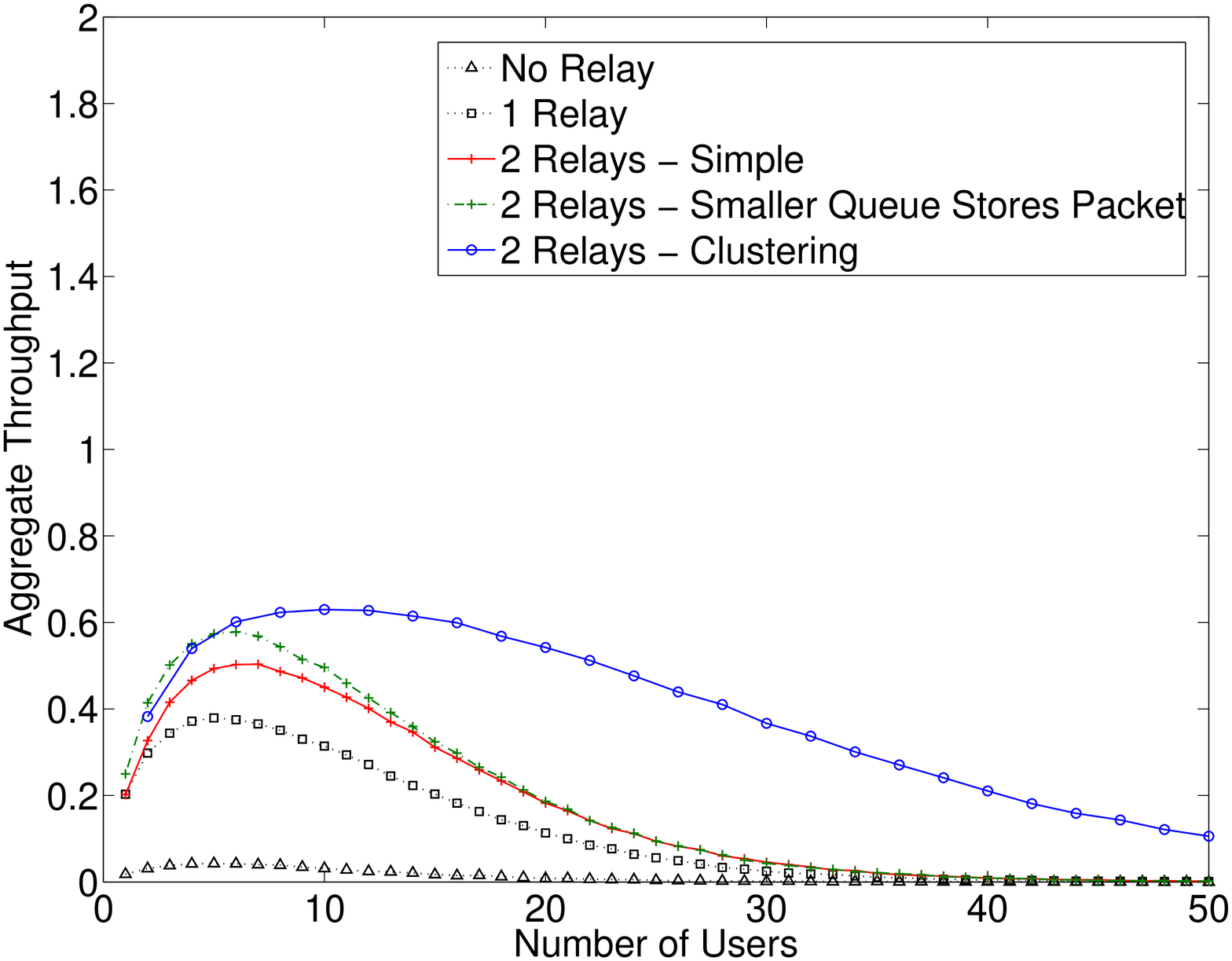}
\caption{$\gamma = 2.5$}\label{subfig:MPRg2.2agg}
\end{subfigure}
\caption{Aggregate throughput (in packets per slot) vs. the
number of users, for different SINR threshold $\gamma$ values.}
\label{fig:MPRagg}
\end{figure}

\begin{figure}[!t]
\centering
\begin{subfigure}[b]{0.45\textwidth}
\includegraphics[width=\textwidth]{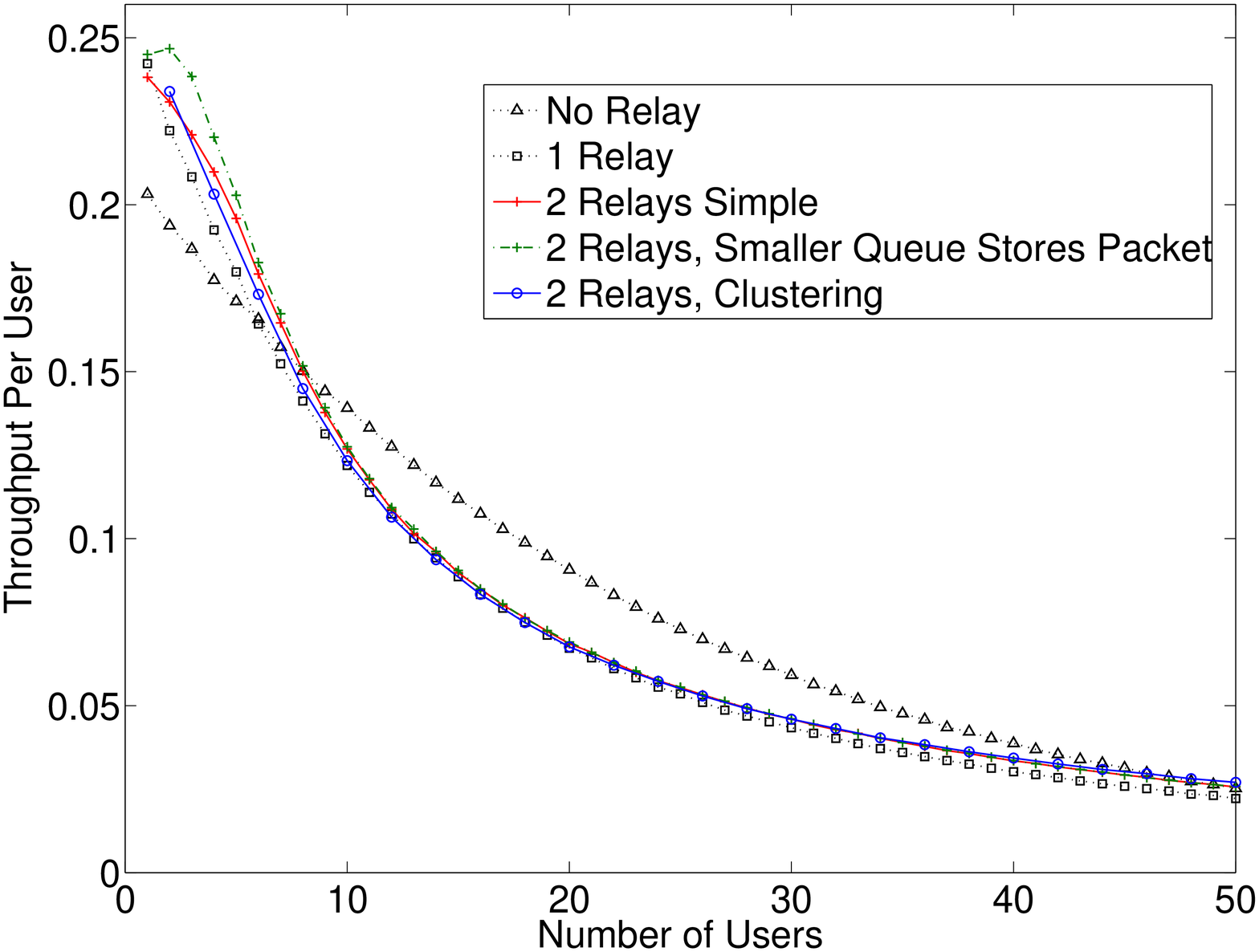}
\caption{$\gamma = 0.2$}\label{subfig:MPRg0.2per}
\end{subfigure}
\begin{subfigure}[b]{0.45\textwidth}
\includegraphics[width=\textwidth]{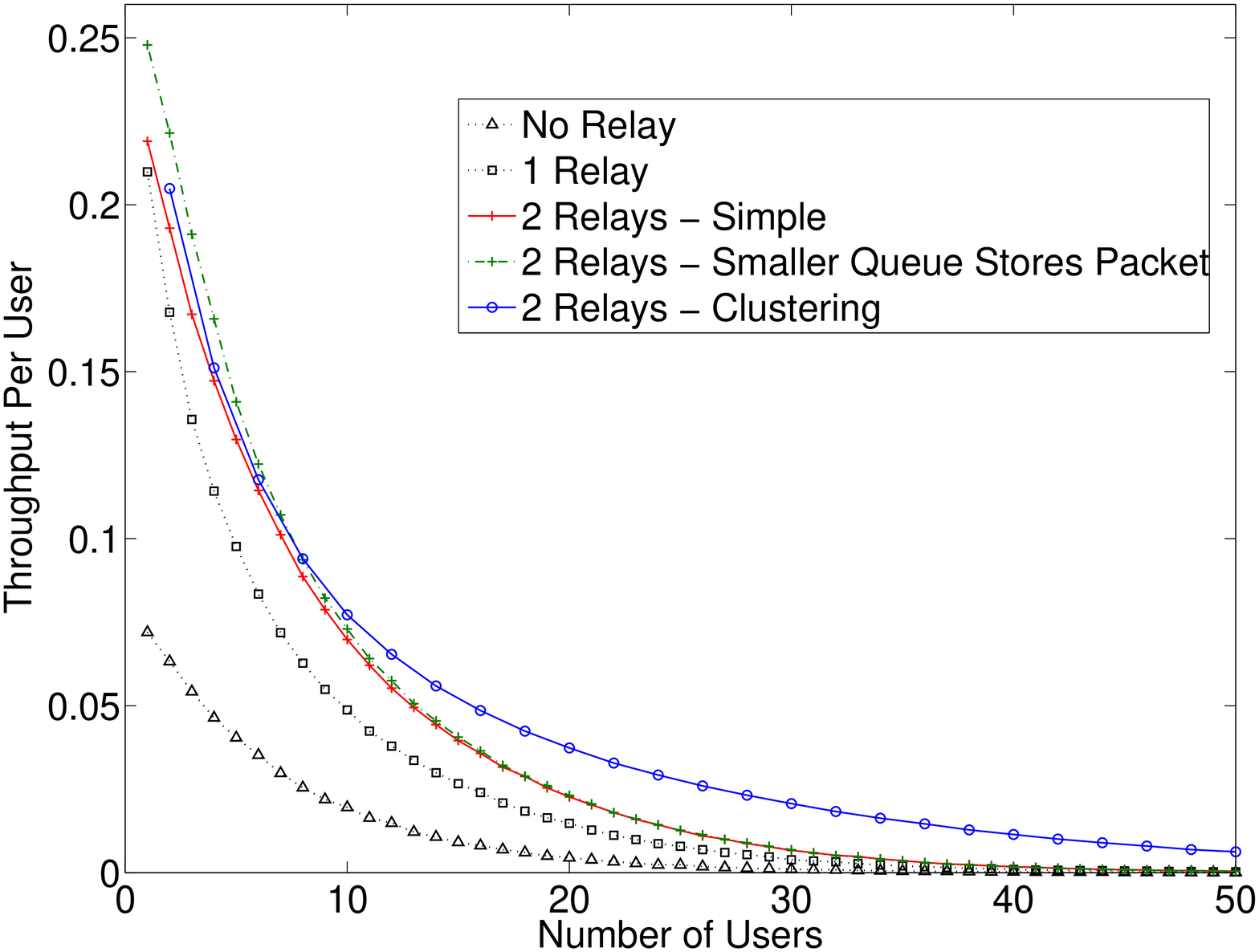}
\caption{$\gamma = 1.2$}\label{subfig:MPRg1.2per}
\end{subfigure}
\begin{subfigure}[b]{0.45\textwidth}
\includegraphics[width=\textwidth]{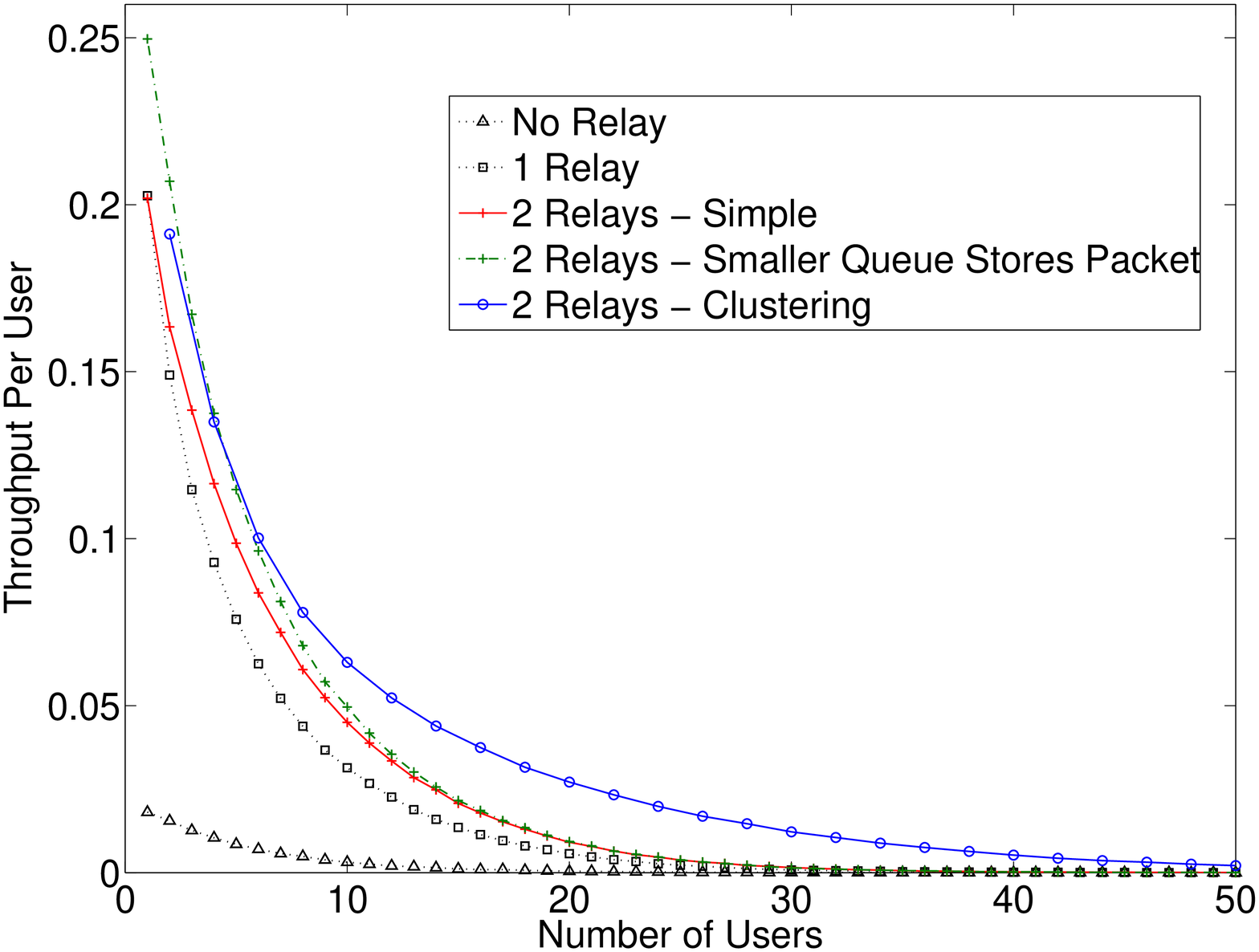}
\caption{$\gamma = 2.5$}\label{subfig:MPRg2.2per}
\end{subfigure}
\caption{Throughput per user (in packets per slot) vs. the number
of users, for different SINR threshold $\gamma$ values.}
\label{fig:MPRPU}
\end{figure}

For a low SINR threshold value ($\gamma=0.2$), the aggregate
throughput (Fig. \ref{subfig:MPRg0.2agg}) and the throughput per
user (Fig. \ref{subfig:MPRg0.2per}) obtained from the system with
two relays are consistently, albeit slightly, higher compared to
that of one relay. For the aggregate throughput, this gain
increases as more users are inserted in the system. However, it
is noteworthy that without relays, the system outperforms those
with relay(s) regardless of clustering or forwarding strategy,
for 8 users and up to a little over 45 users. For more than that,
the performance (always in terms of aggregate and throughput per
user) by the system with two relays is higher compared to the
system with no relay and increases as more users are inserted in
the system. Enabling clustering of users in the two relays starts
providing clear benefits over 30 users, by an approximate 15\%.

Unlike the limited performance gains observed under $\gamma =
0.2$ with the higher SINR thresholds (Figs.
\ref{subfig:MPRg1.2agg} and \ref{subfig:MPRg1.2per} for $\gamma =
1.2$, and  \ref{subfig:MPRg2.2agg} and \ref{subfig:MPRg2.2per}
for $\gamma = 2.5$) we observe that the system with two relays
offers significant advantage compared to the networks without or
with one relay. This is expected since for higher values of
$\gamma$ the relays, having better channel conditions than the
destination, receive a larger percentage of the transmitted
packets in their queues to forward to the destination. Regarding
the forwarding strategy (be it ``Simple'' or the ``Smaller Queue
Stores Packet''), across the threshold value there is a common
trend that for a few users (less than 10 in low $\gamma$, while
less than almost 20 in higher $\gamma$ values) the latter
strategy outperforms the simple one. Furthermore, with two relays
and clustering, in higher SINR thresholds significant advantages
are observed, for over about 10 users. This is again expected
because in each cluster the users interfere with only half the
users of the system in to successfully reach the corresponding
relay (the interference caused in each cluster's relay, by the
users of the other cluster is almost negligible due to the
distance and channel properties).

Note that, the behavior trends across the four relaying schemes
remains stable above $\gamma = 1.2$, in both aggregate throughput
and throughput per user. With this in mind, taking a system
design perspective, one can finally point out that given the link
characteristics and the transmission probabilities, all relaying
schemes reach a maximum aggregate throughput, thus depending on
the number a system is expected to serve, the network designer
can deploy the most appropriate relaying scheme.

\subsection{Average Queue Size}
\label{subsec:Q}
In cooperative systems with relays, a key parameter to be taken
into account is the queue size of the relays. It is important not
only to keep the queues of the relays stable but also to keep
their sizes as low as possible to limit delays.

\begin{figure}[!t]
\centering
\begin{subfigure}[b]{0.45\textwidth}
\includegraphics[width=\textwidth]{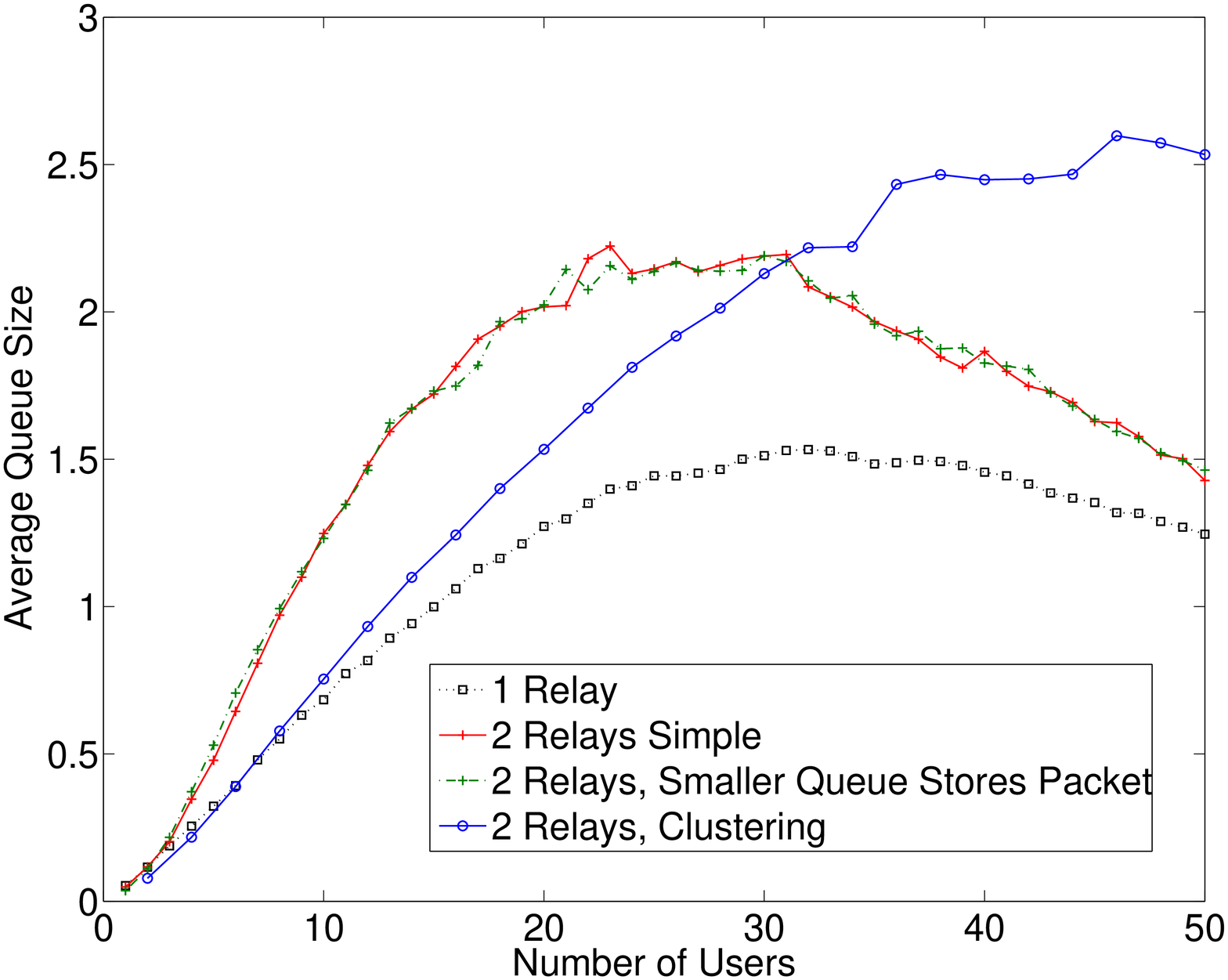}
\caption{$\gamma = 0.2$}\label{subfig:MPRg0.2Q}
\end{subfigure}
\begin{subfigure}[b]{0.45\textwidth}
\includegraphics[width=\textwidth]{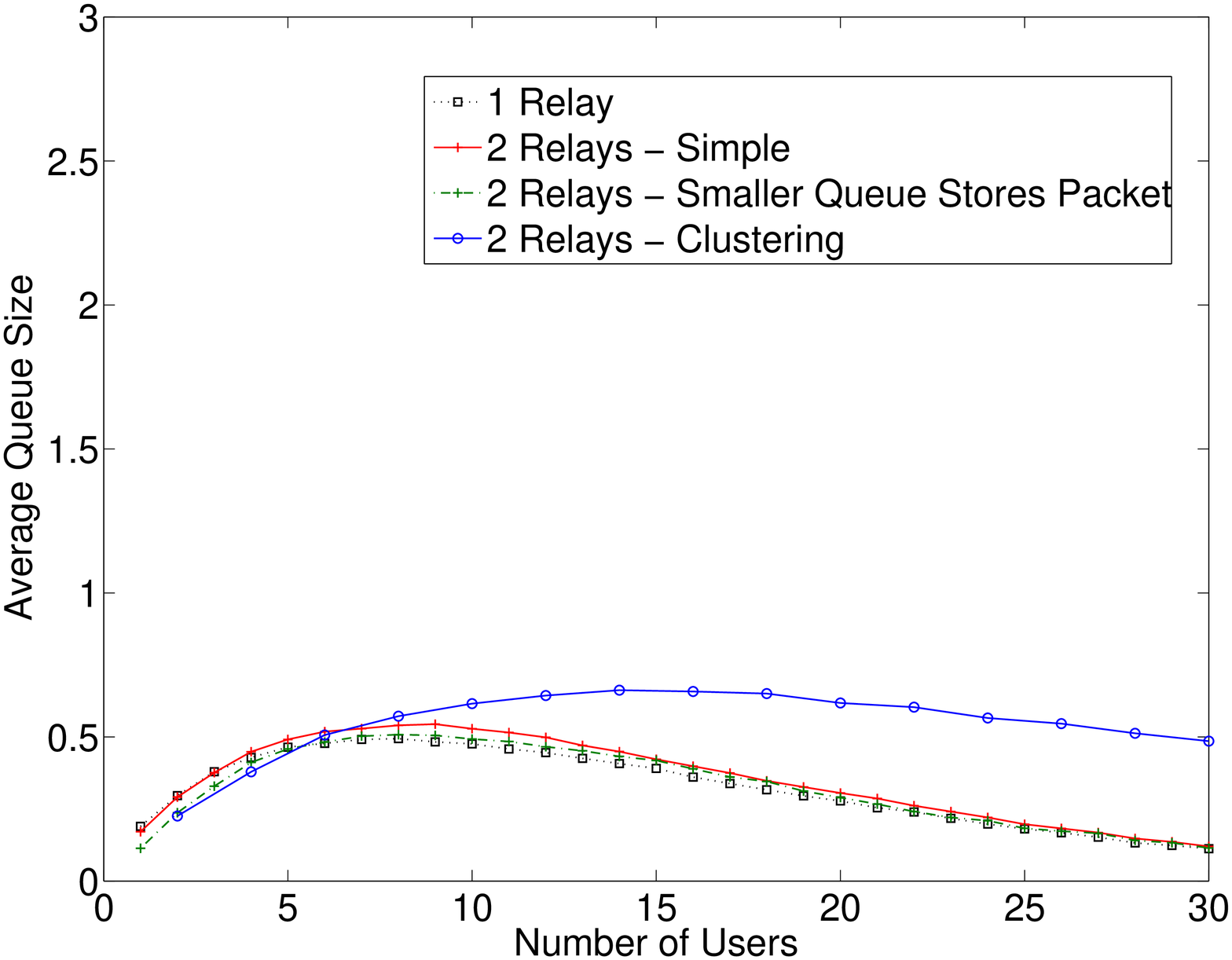}
\caption{$\gamma = 1.2$}\label{subfig:MPRg1.2Q}
\end{subfigure}
\begin{subfigure}[b]{0.45\textwidth}
\includegraphics[width=\textwidth]{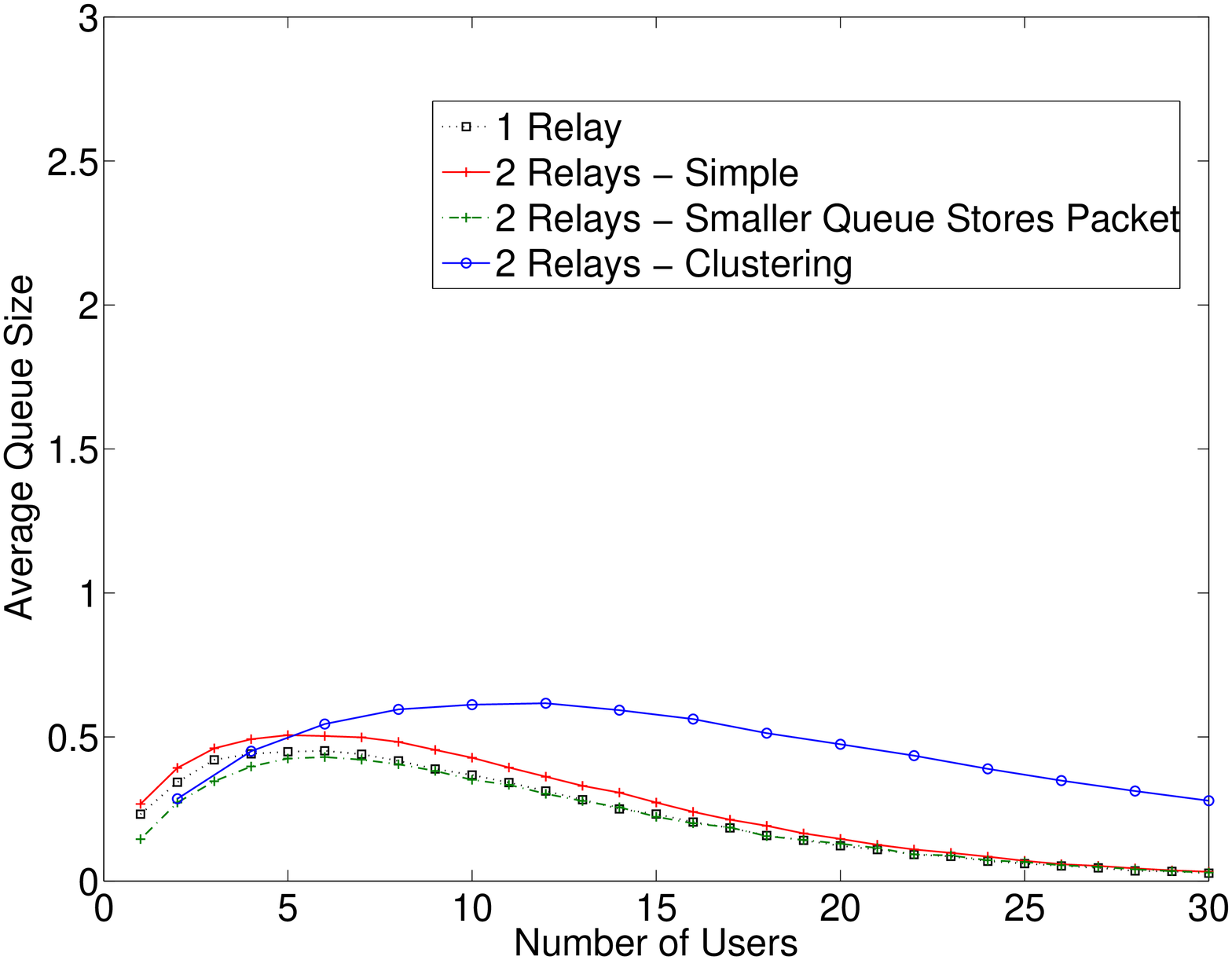}
\caption{$\gamma = 2.5$}\label{subfig:MPRg2.2Q}
\end{subfigure}
\caption{Average relay queue size (in packets) vs. number of
users for different SINR $\gamma$ values in a system with relays.}
\label{fig:Q}
\end{figure}

The plots of Fig. \ref{fig:Q} present the average queue size (in
packets) versus the number of users for the systems with one and
two relays studied in previous sections for $\gamma=0.2$ up to
$\gamma=2.5$. For the systems with two relays only the average
queue size of the one relay is presented (the average queue size
of the second is almost equal because we assume that we have
symmetric users in the systems).

The plots in the figure show that the average queue size of the
clustered system is higher compared to the other systems for a
number of users that reduces with the increase of the threshold
$\gamma$. This is expected because as each relay serves half the
users of the system, the interference between them in the
corresponding relay is lower and more simultaneous transmissions
to a relay may be successful in a time slot. In that way, the two
relays receive more packets resulting in higher queue sizes. It
is interesting though to note that the maximum average queue size
of the system is below one packet, for the two  higher threshold
values (about 0.65 packets for $\gamma=1.2$ and 0.6 for
$\gamma=2.5$). Moreover, in these cases the average queue sizes
of the other three systems tend to become equal with over 25
users.

\subsection{Average per Packet Delay}
\label{subsec:DL}

\begin{figure}[!t]
\centering
\begin{subfigure}[b]{0.45\textwidth}
\includegraphics[width=\textwidth]{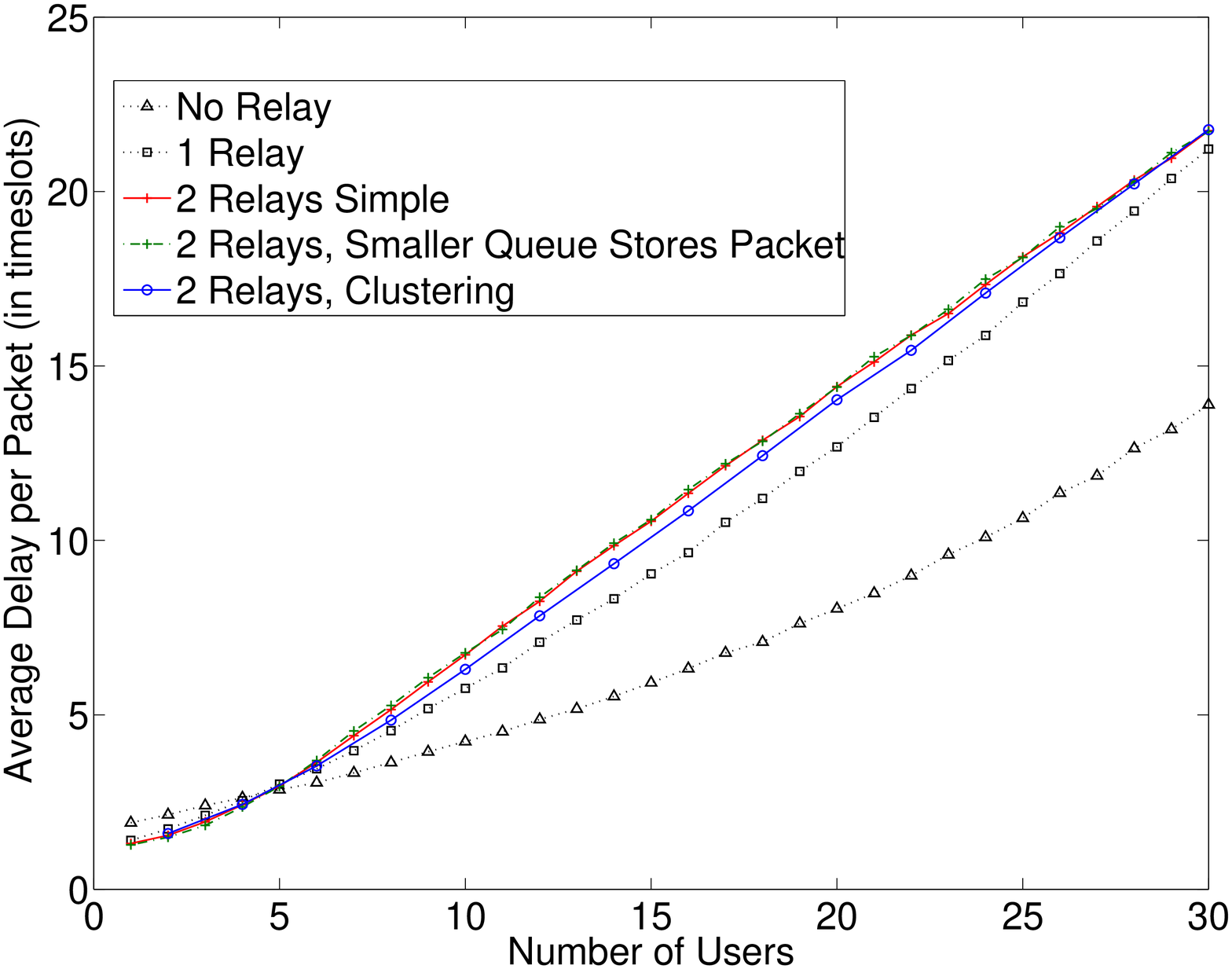}
\caption{$\gamma = 0.2$}\label{subfig:MPRg0.2DL}
\end{subfigure}
\begin{subfigure}[b]{0.45\textwidth}
\includegraphics[width=\textwidth]{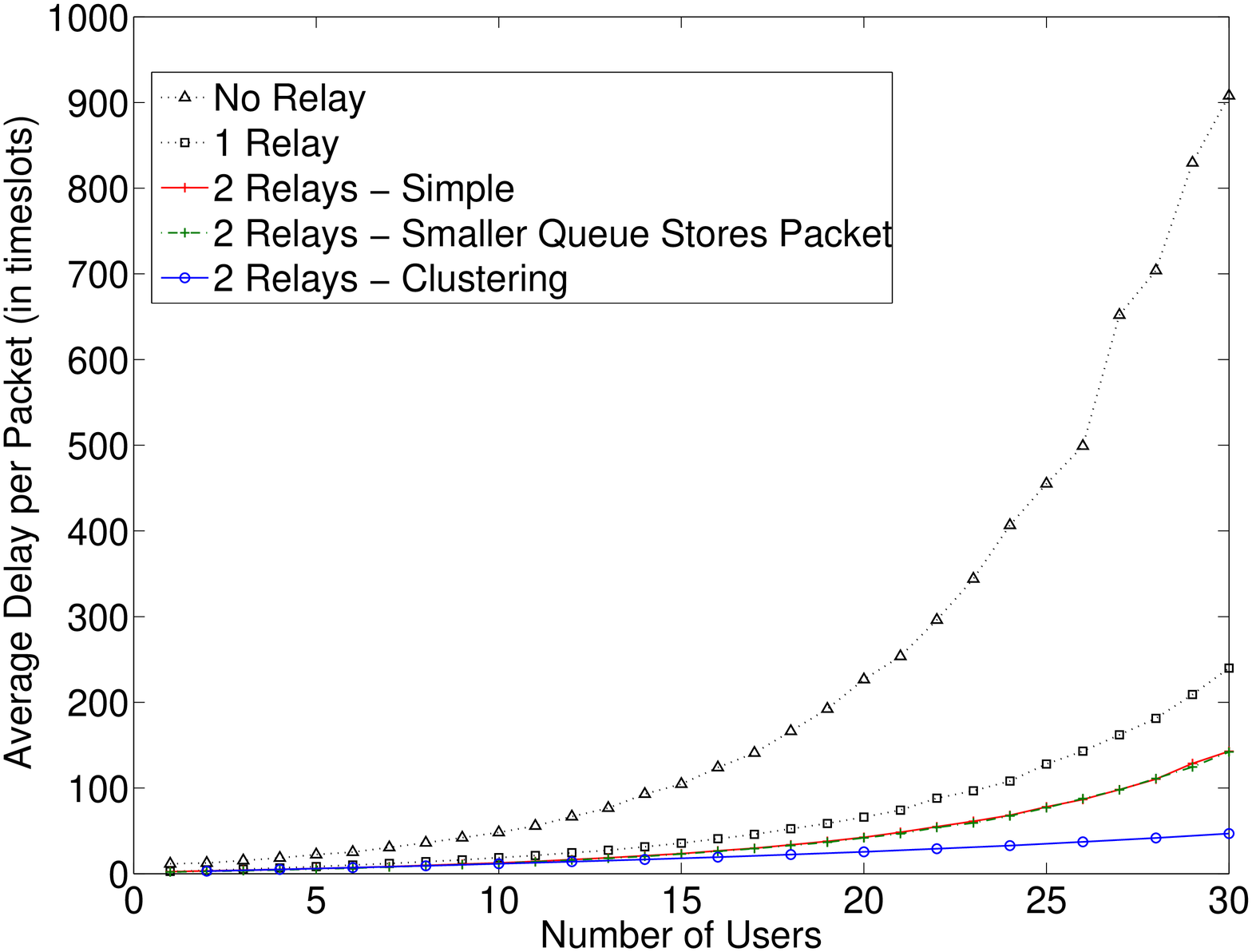}
\caption{$\gamma = 1.2$}\label{subfig:MPRg1.2DL}
\end{subfigure}
\begin{subfigure}[b]{0.45\textwidth}
\includegraphics[width=\textwidth]{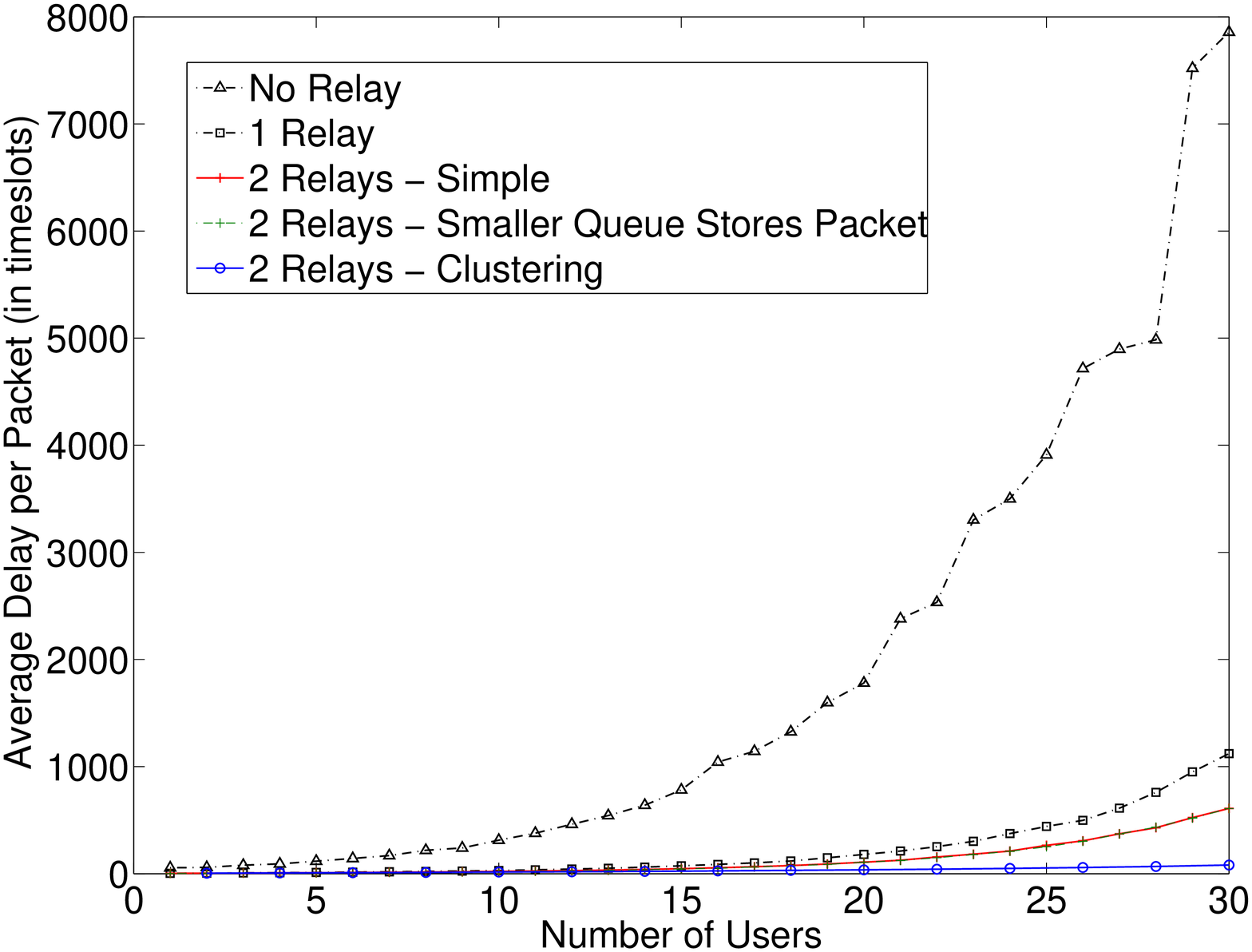}
\caption{$\gamma = 2.5$}\label{subfig:MPRg2.2DL}
\end{subfigure}
\caption{Average per packet delay (in timeslots) vs. number of
users for different SINR $\gamma$ values in a system with and without
relays.}
\label{fig:DL}
\end{figure}

Fig. \ref{fig:DL} presents the average delay per packet (counted
in timeslots) versus the number of users, for $\gamma=0.2$ up to
$\gamma=2.5$ for the five relaying cases presented in the
previous sections. As expected, with more users inserting packets
into the system, the average delay per packet increases due to
the increased interference. Figs. \ref{subfig:MPRg1.2DL} and \ref{subfig:MPRg2.2DL} 
show that the systems with two relays provide less average delay, compared to the systems
with one and no relay, when the number of users is larger than
10. Specifically, for 30 users and $\gamma=1.2$ in Fig.
\ref{subfig:MPRg1.2DL}, the clustered system offers the lowest
average delay per packet and it is interesting to note that
whereas its value increases as the number of users also
increases, it does not exceed 50 timeslots, while for the two
other cases with two relays its value is about $150$ timeslots
and for the one relay  240 timeslots and for no relay 900
timeslots. We can make similar observations from Fig.
\ref{subfig:MPRg2.2DL}.

Furthermore, the average delay per packet for all the cases
except the clustered one increases excessively, for more than 30
users when $\gamma=1.2$ and 25 users when $\gamma=2.5$
respectively. Also, due to the fact that the aggregate throughput
is fairly low and tends to zero as the number of users tend to 50
(see Figs.~ \ref{subfig:MPRg1.2agg} and
~\ref{subfig:MPRg2.2agg}), there are not enough samples in order
to make an accurate calculation of the average delay per packet.
However, the simulation showed that the average delay per packet
obtained from the clustered system for $50$ users and
$\gamma=1.2$ is no more than $160$ timeslots and for $50$ users
and $\gamma=2.5$ it is no more than $460$ timeslots.

%\vspace{-5mm}

%%%%%%%%%%%%%%%%%%%%%%%%%%%%%%%%%%%%%%%%%%%%%%%%%%%%%%%%%%%%%%%%%%%%%%%%%%%%%%%%%%%%%%%%%%%%%%%%%%%%%%%%%
\section{Conclusions}
\label{sec:conclusions}

In this paper, we examined the potential gains of utilizing two
relay nodes to aid the communication of a number of users to a
common destination by re-transmitting (when necessary) their
packets. Under the classic collision channel model we obtained
analytical expressions for the arrival and service rates of the
queues of the two relays and also the stability conditions. We
further showed that the two relays are free to choose their
transmission probabilities independently from each other,
provided that these are greater from some minimum values which
guarantee the stability of their queues. Employing multi-packet
reception made the system intractable, so we conducted a thorough
simulation study.

Under both models, we presented a user clustering scenario where
the users are divided into two groups, each served by one relay
and studied the impact of clustering on the per user and
aggregate throughput.
Although the insertion of a second relay in a system generally
does not offer significantly higher throughput per user in
comparison to a system with one relay, the clustered system
offers impressive performance gains, in terms of throughput, for
large numbers of served users.

Furthermore under the MPR model, we presented two relaying
strategies: a simple one, where if both relays receive the same
packet they both store it and forward it to the destination, and
the Smaller Queue Stores Packet, in which the relay with the
smaller queue becomes responsible for forwarding it to the
destination. The second strategy offers higher aggregate and
throughput per user compared to the first, for limited numbers of
users.

These results could be used, for example in cellular and sensor
networks, to identify the number of required relays to be
deployed and allocate the users among  relays. Future extensions
of this work can include users with non-saturated queues (i.e.
users-sources with external random arrivals) and relays with
their own packets and priorities for the users. Other interesting
extensions consist of relays which are capable of transmitting
and receiving at the same time and the investigation of energy
consumption in the total network and in particular at the relay
nodes.
%%%%%%%%%%%%%%%%%%%%%%%%%%%%%%%%%%%%%%%%%%%%%%%%%%%%%%%%%%%%%%%%%%%%%%%%%%%%%%%%%%%%%%%%%%%%%%%%%%%%%%%%%

\bibliographystyle{ieeetr}
\bibliography{bibliography}

\end{document}